\documentclass[prx, twocolumn]{revtex4-2}

\usepackage{placeins}
\usepackage{float}
\usepackage{amsfonts}
\usepackage{amsmath}
\usepackage{amssymb}
\usepackage{graphicx}
\usepackage{epstopdf}
\usepackage{color}
\usepackage{lipsum}
\usepackage{titlesec}
\newcommand*{\justifyheading}{\raggedright}
\titleformat{\section}
  {\normalfont\large\bfseries\justifyheading}{\thesection}{0.2em}{}
\titleformat{\subsection}
  {\normalfont\large\bfseries\justifyheading}{\thesubsection}{0.2em}{}
\titleformat{\subsubsection}
  {\normalfont\bfseries\justifyheading}{\thesubsubsection}{0.2em}{}

\input epsf.sty

\begin{document}

\title{Quantum Spin Liquid phases in Kitaev Materials}
\author{Po-Hao Chou$^{1,4}$, Chung-Yu Mou$^{2,3,*}$, Chung-Hou Chung$^{1}$, and Sungkit Yip$^{4,5}$}
\affiliation{$^{1}$Electrophysics Department, National Yang Ming Chiao Tung University, Hsinchu, Taiwan 300, R.O.C.}
\affiliation{$^{2}$Center for Quantum  Science and Technology, National Tsing Hua University, Hsinchu, Taiwan 300, R.O.C.}
\affiliation{$^{3}$Department of Physics, National Tsing-Hua University, Hsinchu, Taiwan 300, R.O.C}
\affiliation{$^{4}$Institute of Physics, Academia Sinica, Taipei 115, Taiwan, R.O.C.}
\affiliation{$^{5}$Institute of Atomic and Molecular Sciences, Academia Sinica, Taipei 106, Taiwan, R.O.C.}
\thanks{Corresponding author. Email: mou@phys.nthu.edu.tw}


\begin{abstract}
We develop a gauge-invariant renormalized mean-field theory (RMFT) to reliably find the quantum spin liquid (QSL) states and their field response for realistic Kitaev materials under strong magnetic fields and described by the generalized Kitaev $J$-$K$-$\Gamma$-$\Gamma'$ model. Remarkably, while our RMFT reproduces previous results based on using more complicated numerical methods, it also predicts several new stable QSL states. In particular, since Kitaev spin liquid (KSL) is no longer a saddle point solution, a new exotic 2-cone state distinct from the KSL, is found to describe experimental observations well, and hence should be the candidate state realized in the Kitaev material, $\alpha$-RuCl$_3$. We further explore the mechanism for the suppression of the observed thermal Hall conductivity at low temperatures within the fermionic framework, and show that the polar-angle dependence of the fermionic gap can distinguish the found 2-cone state from the KSL state in further experiments.
\end{abstract}
\maketitle

\noindent The quantum spin liquid (QSL) is a state of quantum spins without spin ordering but spins are highly entangled\cite{Balent}.
The original idea that interacting spins may form a liquid phase was first proposed by P. W. Anderson\cite{Anderson1}. It was later brought up by Anderson to construct the parent state that fosters superconductivity for doped cuprate superconductors\cite{Anderson2}. The realization of Anderson's idea  has been pursued in many interacting spin systems. In particular, the quantum spin liquid phase has been demonstrated to be the ground state for a number of theoretical models. Notably, the Kitaev model on honeycomb lattice is an exactly solvable model with the quantum spin liquid phase being the ground state\cite{Kitaev1}.

Being an exactly solvable model,  the Kitaev model provides a platform to understand gapped and gapless quantum spin-liquid (QSL) states with distinct anyon statistics. The gapless Kitaev spin-liquid (KSL) state possesses two kinds of basic excitation, cone-like gapless fermionic excitation formed
by Majorana fermions and gapped excitation formed by flux excitations, termed as vison. Under weak magnetic fields, the Majorana-fermion excitations
open up a gap with their bands possessing finite Chern number
$\nu = \pm 1$\cite{Kitaev1,Nasu1,Gohlke1,Zhang}.
The associated Majorana edge states give
rise to half quantized thermal Hall conductivity\cite{Kitaev1,Nasu2}
which is claimed to be observed in some
experiments\cite{Matsuda_Nat,Matsuda_Sci,Matsuda_PRB}. The gapped visons together with Majorana bound states
at their core obey non-Abelian statistics
\cite{Kitaev1,Gohlke1,Hickey}, thus providing
a potential platform for topological quantum computation.

On experimental side,  due to the unusual spin-spin interaction involved, it is difficult to realize the  Kitaev model in pure spin systems. Instead, it was pointed out that super-exchange interactions of pseudospins formed by the combination of orbital and spin degree of freedom for d electrons may realize the required spin-spin interaction\cite{compass1,compass2}. However, in real systems, other exchange interactions are also present, the resulting model usually contains many other spin-spin interactions. The typical model that includes the Kitaev spin-spin interaction is the so-called the generalized Kitaev $J\mbox{-}K\mbox{-}\Gamma\mbox{-}\Gamma^{\prime}$ model in which $K$ is the strength of the Kitaev spin-spin interaction, $J$ is the isotropic super-exchange interaction, $\Gamma$ and $\Gamma'$ are interactions of off-diagonal components of pseudospins\cite{Kee2,Kee3,Janssen,Pontus,Katukuri,Yamaji}.  Material candidates for realizing the model are mainly in iridium oxides, ruthenium and cobalt compounds\cite{Takagi,Chaebin,Lin,Weiliang}.

Recently, a key feature of  QSL was observed in the candidate Kitaev material $\alpha$-$\mbox{RuCl}_3$.
By applying magnetic fields, it is found that the antiferromagnetic phase in $\alpha$-$\mbox{RuCl}_3$ evolves into
a phase with half-integer quantized thermal Hall conductivity in some samples at a given temperature range\cite{Matsuda_Nat,Matsuda_Sci,Matsuda_PRB}.
However, some other experiments reported non-quantized thermal Hall conductivity at low temperatures\cite{Czajka_1,Bruin,Czajka_2}.
It is also suggested that phonons make the major contribution to the thermal Hall conductivity\cite{Lefran}.
Although it is still under debate on whether this phase is a spin-liquid phase and on
the bosonic or fermionic origin of the observed thermal Hall phenomenon\cite{Czajka_1,Bruin,Czajka_2,Kee1},
the field-angle-dependent specific heat measurement seems to support fermionic scenario, as the sign-changing of thermal Hall conductivity
and the fitted band gap seem to agree with predictions by the KSL under weak field\cite{Tanaka,Imamura,Hwang}.
Theoretically, these predictions are based on features of original KSL.
As indicated in the above, however, except for the dominant ferromagnetic Kitaev interaction $K$,
the effective spin model for  $\alpha$-$\mbox{RuCl}_3$ contains a large first-neighbor off-diagonal $\Gamma$ interaction,
possible small $J$ and $\Gamma'$ interactions, and third-neighbor Heisenberg interaction\cite{Kee2,Kee3,Janssen,Pontus,Katukuri,Yamaji}.
It is known that the KSL is unstable with the presence of these extra interactions\cite{Gohlke2,Kee4,Wang_PRL,Lee,toronto,Yilmaz}.
Therefore, it calls for a theoretical explanation of the observed phenomenon.

On the theoretical side, to study fermionic spin liquid state, a widely-used method is first employing the Kitaev Majorana-fermion decomposition of spin to construct quadratic
mean-field (MF) Hamiltonian, and then rewriting the Hamiltonian back to standard Schwinger-Fermion form. With the general variational parameters of MF Hamiltonian
provided by projective symmetry group (PSG) analysis\cite{PSG_Wen,PSG_Burnell,PSG_You}, the corresponding Gutzwiller projected variational wave-function is used to optimize the variational energy by variational Monte Carlo (VMC) method\cite{Wang_PRL,Wang_PRB,Jiang_PRL}. Such a mathematical procedure, however, may become cumbersome once one wants to study the effect of external magnetic fields in arbitrary orientations as many relevant terms arise due to the broken symmetries of the Hamiltonian. The VMC method is also limited by the system size and would not be able to find results in the thermodynamic limit. In addition, though  the Kitaev Majorana-fermion decomposition works well in the Kitaev model, it suffers from the problem of being gauge non-invariant in the general spin interaction model.
As  we will show, the generalized Kitaev $J\mbox{-}K\mbox{-}\Gamma\mbox{-}\Gamma^{\prime}$ model would unphysically yield the same QSL solution in the mean-field level for both positive and negative parameters, ($\pm J$, $\pm K$ $\pm \Gamma$ $\pm \Gamma^{\prime}$)(as an example, one may consider the $J\rightarrow\infty$ limit). Therefore, it also calls for a new theoretical approach to account for the observed phenomenon.

To overcome the above mentioned difficulty encountered in VMC, in this work, we apply the renormalized mean-field theory (RMFT)
to find the quantum spin liquid states in the presence of magnetic fields. The renormalized mean-field theory was a reliable method introduced to find
correlated electronic phases in the Hubbard model after the high Tc cuprate superconductors were discovered\cite{RMFT1, RMFT2, RMFT3,RMFT4}.
The theory has achieved great success for systems being spin-SU(2) invariant and  was originally aimed to replace the rigorous approach based on VMC method by a simplified approximation in which effects of the Gutzwiller projection
are approximated by classical weights —namely, by the renormalized factors (such as $g_J$ and $g_m$ defined in in the following sections)
for the coupling constants of various operators in the Hamiltonian. It was later generalized to include local order parameters and correlations in the renormalization factor\cite{RMFT5,RMFT6}. Here we shall further generalize RMFT by computing the renormalized factor directly and extend RMFT to systems without spin-SU(2) symmetry. For this purpose, we first employ the gauge invariant decomposition of the spin operator\cite{MajDecom1,MajDecom2}.
Adopting this decomposition, a gauge invariant MF theory that includes quantum averages over all allowed local clusters of operators
is derived. By comparing with the exact solution of the Kitaev model, we find that if the coupling constant is renormalized by
a renormalized factor $g_J$, the MF solution can reproduce all results of the ground state that  is based on the exact solution.
We rigorously prove that $g_J$ is exactly equal to $4$ for the Kitaev model. The Kitaev's exact solution provides an excellent example
of the renormalized mean-field theory in which we have generalized it by replacing the classical weighting factors with factors determined by
quantum averages over local clusters of operators.
When the interaction $J$,$\Gamma$,$\Gamma^{\prime}$ or magnetic moments are present, the $g_J$ factor needs to be replaced by a bond-dependent
tensor $g^{\alpha\alpha^{\prime}}_{J,ij}$, and the magnetic moments $m^{\alpha}_i$ also acquire an approximated renormalized factor $g^{\alpha}_{m,i}$.
The resulting RMFT reproduces several results previously obtained by other numerical methods, such as density matrix renormalization group(DMRG) and VMC study of
Kitaev model under field \cite{RMFT_compare,Jiang_PRL}, and in particular, reproduces the $8$-,$14$-,$20$-cone quantum spin-liquid (QSL)
states and the band Chern number under fields in the  VMC study of $K\mbox{-}\Gamma$ model \cite{Wang_PRL,Wang_PRB}.
Therefore, our RMFT approach provides an economic way in the thermodynamic limit to investigate the MF spinon band properties and the vortices structure in $J\mbox{-}K\mbox{-}\Gamma\mbox{-}\Gamma^{\prime}$ model under field in arbitrary orientations. Here the vortices structure refers to the vortices state generalized from the exact solution of Kitaev model\cite{Kitaev1}.

Note that a recent review article lists various types of ground states that were predicted by various numerical methods on the $K$-$\Gamma$ model\cite{Perkin}. We would like to point it out that even though these various advanced numerical methods, including DMRG and tensor network approaches, have proposed a number of ground states for $K$-$\Gamma$ model, a definitive conclusion still remains elusive.
Furthermore, to our knowledge, these methods have not yet been able to reproduce key experimentally observed spin-liquid features under experimental magnetic field orientation, such as chiral edge channels and the six-fold angular dependence of the fermionic gap. While some of the methods capture a spin-liquid ground state\cite{Kee4,Perkin}, they inherently struggle to describe such features due to their $2\times2$ spin representation and are limited to system sizes or geometry.  For instance, the $K$-$\Gamma$ spin-liquid reported in Ref. \onlinecite{Kee4}, where the Chern number under field is lacking. In addition, they do not report on other characterization of the phases, e.g., excitations etc.
In addition, to describe realistic Kitaev materials, one should adopt the generalized Kitaev $J$-$K$-$\Gamma$-$\Gamma'$ model instead
of the $K$-$\Gamma$ model compared in  Ref. \onlinecite{Perkin}.
Therefore, in the following, we direct our efforts toward identifying the robust spin-liquid phase within the $J$-$K$-$\Gamma$-$\Gamma'$ model under experimental magnetic field orientation, with all parameters chosen within the realistic material range. We then examine the magnetic response of these states to compare with experimental observation\cite{Matsuda_Nat,Matsuda_Sci,Matsuda_PRB,Tanaka,Imamura}. More detailed descriptions of our results are shown in below.

Based on this new RMFT approach, we obtain several new results that were not found before. The main results from our RMFT are as follows:
Under zero field and $K=-1, \Gamma\gtrsim0.25$ there are several new stable QSL states: $Z_2$ 2-, 4-, 8-, 14-, 20-, 26-cone QSL states,
$SU(2)$ 16-, 32-cone QSL states and other high number cones QSL states.
Some of these states remain robust in the parameter regime with large $\Gamma(>0.5)$ and small $|\Gamma'|$, $|J|(<0.1)$, which corresponds to the typical range for
$\alpha$-RuCl$_3$. Since their energies per site are very close to each other with difference $\lesssim 0.02|K|$(2$\sim$4 $K$), we futher examine their low field band structure, Chern numbers, and angle-dependent gap under magnetic fields to compare with the experimental results.
Remarkably, we find that only the $Z_2$ 2-cone state yields the Chern number
$\nu=\pm 1$ at low fields, and at the same time it gives the same azimuthal angle-dependent gap observed by the experiment under magnetic fields.
Since this 2-cone state possesses negative Wilson loop value, it is distinct from the Kitaev spin-liquid state.
When the applying field is in the $\hat{a}$-$\hat{b}$ plane, this 2-cone state exhibits opposite Chern numbers in comparison to KSL. By further examining
the ground state degeneracies, we also find this 2-cone state is a topological QSL with long-range entanglement\cite{TP_Wen}. Since
the Kitaev spin liquid is not stable in the presence of large $J$, $\Gamma$, and $\Gamma^{\prime}$\cite{Gohlke2,Kee4,Wang_PRL,Lee,toronto,Yilmaz},  it implies that
what the experiments observed is the 2-cone state obtained in our RMFT.  We further illustrate that this 2-cone state can be distinguished from the Kitaev
spin-liquid state by looking into the angle-dependent gap in the polar angle (in contrast to the azimuthal angle currently observed in experiment).
Finally, we discuss possible mechanism for the suppression of the observed thermal Hall conductivity at low temperatures within the  Majorana-fermion framework and conclude.

\section*{Results}
\subsection*{Theoretical model and Gauge Invariant Decomposition of Spin}
\noindent We start from the generalized Kitaev $J\mbox{-}K\mbox{-}\Gamma\mbox{-}\Gamma^{\prime}$ model on the honeycomb lattice\cite{Janssen,Pontus},
which under magnetic field  $\vec{h}$, is given by
\begin{eqnarray} \label{H0}
H =\sum_{\langle ij \rangle \in \gamma_l} J \vec{S}_i\cdot\vec{S}_j+ K {S}^{\gamma}_i {S}^{\gamma}_j
+\Gamma ({S}^{\alpha}_i {S}^{\beta}_j+{S}^{\beta}_i {S}^{\alpha}_j) \nonumber \\
+\Gamma^{\prime}( {S}^{\gamma}_i {S}^{\alpha}_j+ {S}^{\gamma}_i {S}^{\beta}_j
+{S}^{\alpha}_i {S}^{\gamma}_j+{S}^{\beta}_i {S}^{\gamma}_j)-\vec{h}\cdot\sum_{i}\vec{ {S}}_{i}.
\end{eqnarray}

Here $\vec{S}_j$  is not a real spin operator and represents pseudo spin-1/2 operators at site $j$, $\gamma_l$ denotes $x_l$, $y_l$ or $z_l$ bonds shown in Fig. 1 (a) with $(\alpha,\beta,\gamma)$ being the cyclic permutation of the indices $(x,y,z)$, $K$ is the strength of  the Kitaev spin-spin interaction,
$J$ is the isotropic super-exchange interaction, and $\Gamma$ and $\Gamma'$ are interactions of off-diagonal components of pseudospins. The  Hamiltonian can  be cast in a more concise form  as
\begin{eqnarray} \label{H1}
H=\sum_{\langle ij \rangle, \alpha\alpha^{\prime}} J^{\alpha\alpha^{\prime}}_{ij} {\sigma}^{\alpha}_{i} {\sigma}^{\alpha^{\prime}}_{j}-\frac{\vec{h}}{2}\cdot\sum_{i}\vec{{\sigma}}_{i},
\end{eqnarray}
where all spin-spin interactions are lumped into (including the $1/4$ factor) an appropriate tensor $J^{\alpha\alpha^{\prime}}_{ij}$ and we have made use of the relation $\vec{S}_j=\frac{1}{2}\vec{\sigma}_j$.

Now we need to decompose the spin operator into Majorana fermions. In the standard second quantization of the spin operator in terms of fermionic operators, $f^{\dagger}_{\alpha}$ and $f_{\alpha}$, we have $\vec{\sigma} =  \sum_{\alpha, \beta} f^{\dagger}_{\alpha} \vec{\sigma}_{\alpha \beta} f_{\beta}$, where $\alpha$ and $\beta$ are indices corresponding to either $\uparrow$ or $\downarrow$. To faithfully represent the spin, a projection to the Fock space with each site being singly occupied by fermions is implemented through gauge fields so that $\vec{\sigma}$ and $H$ are gauge invariant\cite{Wen_Lee}.  From $f^{\dagger}_{\alpha}$ and $f_{\alpha}$, one can form four Majorana fermions as  $b^x =f^{\dagger}_{\uparrow}+f_{\uparrow}$,  $b^y =-i f^{\dagger}_{\uparrow}+if_{\uparrow}$, $b^z =-(f^{\dagger}_{\downarrow}+f_{\downarrow})$,  $b^0 =i f^{\dagger}_{\downarrow}-if_{\downarrow}$. We find that the spin decomposition of Majorana fermions at site $j$ reads\cite{PSG_You,MajDecom2}
\begin{eqnarray} \label{sigma0}
\sigma^{\alpha}_j=\frac{1}{2}(ib^{\alpha}_jb^{0}_j-ib^{\beta}_jb^{\gamma}_j),
\end{eqnarray}
where  $\alpha$, $\beta$, and $\gamma$ are cyclic permutation of $x$, $y$, and $z$. Note that in the Kitaev Majorana-fermion decomposition of spin, the spin operator is expressed as $\tilde{\sigma}^{\alpha} = i b^{\alpha} b^{0}$ with the constraint $b^xb^yb^zb^0=1$, where $\alpha = x$, $y$, or $z$.  For the Kitaev spin-spin interaction $H_K$, as only the same component of spins are involved for each bond $(i,j)$ of the lattice, the associated Majorana bilinear product $b^{\alpha}_i b^{\alpha}_j$ commutes with $H_K$. This reduces the problem to diagonalization of a quadratic Hamiltonian and the problem is thus exactly solvable\cite{Kitaev1}. Clearly, in the presence of $J$, $\Gamma$, or $\Gamma'$, the Kitaev decomposition no longer has the advantage and the Hamiltonian is not exactly solvable. Furthermore, since for general spin-spin interaction on bond $(i,j)$, $\sum_{\alpha \beta} J^{\alpha \beta} \sigma^{\alpha}_i \sigma^{\beta}_j = \sum_{\alpha \beta} J^{\alpha \beta} (i b^{\alpha}_i b^{0}_i) (i b^{\alpha}_j b^{0}_j)$, hence for $-J^{\alpha \beta}$, the minus sign can be absorbed into $b^0_i$ by redefining  a new Majorana fermion as $\tilde{b}^0_i=-b^0_i$, while maintaining the constraint $b^x_ib^y_ib^z_ib^0_i=1$ at the mean-field level if no moment term presents. The Kitaev decomposition would unphysically yield the same solution in the mean-field level for both positive and negative parameters. In addition, since
$\tilde{\sigma}^{\alpha} $ can be expressed as $\sigma^{\alpha}- Q^{\alpha}$ with $Q_i^{\alpha}$ being the pseudospin operator given by $Q_i^{\alpha}= (f^{\dagger}_{i,\uparrow} , f_{i,\downarrow}) \tau^{\alpha} (f_{i,\uparrow} , f^{\dagger}_{i,\downarrow})^T$, it is clear that $Q_i^{\alpha}$ rotates covariantly with the pseudospin SU(2) rotations \cite{Wen_Lee} and hence $\tilde{\sigma}^{\alpha}$ is not gauge invariant.

From above, we conclude by using Eq.(\ref{sigma0}) that the Hamiltonian $H$ is also gauge invariant.
It is thus favorable to use Eq.(3) for the spin decomposition. So far, the consideration is for the exact Hamiltonian.  In practice, one needs to
keep the gauge invariance when approximations are taken.  In this case, it should be kept in mind that
 the average of spin-spin interaction, $\sigma^{\alpha}_i\sigma^{\alpha^{\prime}}_j$,
 is gauge invariant only if all allowed Wick's decompositions are taken
\begin{eqnarray} \label{SS} \notag
&\langle \sigma_i^{\alpha}\sigma_j^{\alpha^{\prime}} \rangle=
\frac{1}{4}\left(-\chi_{ij}^{00}\chi_{ij}^{\alpha\alpha^{\prime}}+\chi_{ij}^{0\alpha^{\prime}}\chi_{ij}^{\alpha0}
+\chi_{ij}^{0\gamma^{\prime}}\chi_{ij}^{\alpha\beta^{\prime}}-\chi_{ij}^{0\beta^{\prime}}\chi_{ij}^{\alpha\gamma^{\prime}}\right. \\
&\left.+\chi_{ij}^{\gamma0}\chi_{ij}^{\beta\alpha^{\prime}}-\chi_{ij}^{\gamma\alpha^{\prime}}\chi_{ij}^{\beta0}
-\chi_{ij}^{\gamma\gamma^{\prime}}\chi_{ij}^{\beta\beta^{\prime}}+\chi_{ij}^{\gamma\beta^{\prime}}\chi_{ij}^{\beta\gamma^{\prime}}\right)+m_i^{\alpha}m_j^{\alpha^{\prime}},  \nonumber
\end{eqnarray}
where $\chi_{ij}^{\mu\mu^{\prime}}=\langle i b^{\mu}_i b^{\mu^{\prime}}_j \rangle$, and $m^{\alpha}_i=\langle \sigma_i^{\alpha} \rangle $.
Based on the gauge-invariant decomposition of spin, we shall derive the gauge-invariant renormalized mean field theory (RMFT), which is done in the section of methods.

\subsection*{Comparison with early numerical works}
\noindent Before presenting results for the generalized Kitaev model, we shall first compare solutions based on our RMFT with results based
on other numerical results for the Kitaev model under field.

In Fig. 2, we compare our results with those based on
methods of density matrix renormalization group (DMRG) and variational Monte Carlo (VMC).  Clearly, good
agreement is obtained for the magnetization curve in Fig. 2 (a) and energy in Fig. 2 (b) of Kitaev model under field.
Furthermore, for some range of $\Gamma$'s, Wang\cite{Wang_PRL} found a 14 cone states,
and therefore we would like to compare with their results in Fig. 2 (c).
In particular, our RMFT results also show the same pattern of
changing for the topological Chern number under the magnetic field along $\hat{c}$ direction, though the exact boundary of change is
different due to the energy comparison between different states. Detailed comparison of the band gap and Chern number are shown
in Supplementary Figure 6. Results in Fig. 2 and Supplementary Figure 6 indicate that this gauge-invariant RMFT is a reliable method.

It is worth to mention, if one uses MF based on Kitaev's decomposition (KMF) in $K$-$\Gamma$ model, one would not get listed stable QSL found by VMC or RMFT, only GKSL survives;
while if one uses MF based on the gauge invariant decomposition but without including the cluster renormalization effect,
for $K<0$ Kitaev model, one would get a negative spin susceptibility, which is clearly wrong.

In Fig. 3, we compare our RMFT results with the well-established results of the $J$-$K$ model\cite{KJmodel}. It is evident that RMFT yields relatively poor energy in the magnetically ordered phases. This discrepancy arises from the fact that the RMFT approach employs the representation in which spinons are deconfined in the sense that the spin operator is broken into two unbinded spinons.
Physically, however, in a magnetically ordered phase or a polarized phase where large magnetic moments are present, spinons are expected to be confined so that two spinons recombines into the spin operator. As a result, fluctuations of the confinement potential (represented by the introduced Lagrange multipliers) cannot be neglected, and treating them merely at the mean-field level is insufficient. On the other hand, most advanced numerical methods are based on representations in terms of spin (which confines two spinons into the spin operator) and it would be more difficult for these methods to find spin-liquid. Note that in the ordered phase, the classical mean-field solution in which $\chi_{ij}$ vanishes is always a solution to our RMFT theory.  However, it does not imply that at zero magnetic field, our RMFT solution coincides with the classical mean field solution. Instead, what we found is that at zero magnetic field, our RMFT solution does not coincide with the classical mean-field solution by the presence of finite $\chi_{ij}$ in the ordered phase. This is because in ordered phase, the presence of finite $\chi_{ij}$ makes the energy of our RMFT solution even lower than that of the classical mean-field solution. Hence even though the classical mean-field solutions in are solutions to our RMFT, they are only local minima and are not the true ground states.

In Fig. 4, we compare our RMFT results with the DMRG results of the $K$-$\Gamma$ model under field\cite{Kee4}.  Noted that the DMRG results were obtained on a two-leg honeycomb strip—i.e., with only two unit cells in one direction—which may lead to the differences in the phase boundaries. An important similarity is that both their study and ours identify a spin-liquid phase in the large Gamma region: their $K\Gamma$ spin-liquid and our 2-cone spin-liquid(see next section), respectively. Moreover, the Wilson loop value W of their $K$-$\Gamma$ spin-liquid is approximately -1/3, while the value of W of our 2-cone spin-liquid is approximately -0.35 (see Table. II).

\subsection*{QSL states with full symmetries}
\noindent The symmetry group for the Hamiltonian in zero field and the corresponding projective symmetry group
are  analyzed in Secs. I and V of the Supplemental Material\cite{SMI}.
Only six irreducible MF parameters are allowed if we require a spin liquid state to
possess full symmetries, they are given by
\begin{eqnarray} \label{chizl}
\chi^{00}_{z_l},\chi^{xx}_{z_l},\chi^{zz}_{z_l},\chi^{0b}_{z_l},\chi^{xy}_{z_l},\chi^{xz}_{z_l}
\end{eqnarray}
on $z_l$-link\cite{SMI}, where $\chi^{0b}_{z_l}\equiv\chi^{0x}_{z_l}=-\chi^{0y}_{z_l}$.

In \cite{Wang_PRL} and \cite{Wang_PRB} they claimed there are two terms (their $\eta_{3,5}$) that couple
$b^0$ to $b^{\mu \ne 0}$. However, one can check that
their sum results in a term that does not couple these two sets of operators.
Hence there is only one linear independent term that couples
$b^0$ to $b^{\mu \ne 0}$, as in ours.
Note that our number of irreducible parameters (six) is the same as
Wang's PRL work\cite{Wang_PRL}.
Though in their later PRB work\cite{Wang_PRB} they claim there are seven
irreducible parameters,
by examining their PRB result, there are only six linear independent terms in it.
Therefore, six is the correct number, also as in ours.

To search the stable quantum spin-liquid states in the accepted parameter region of $\alpha$-RuCl$_3$ with $K=-1$,$\Gamma>0.3$,$|J|,|\Gamma'|<0.1$,
we start from the KSL state with $K=-1$ and $\Gamma=0$. When a small value of  $\Gamma$ is turned on, the KSL state
becomes GKSL (generalized KSL). The GKSL then continuously evolves to one of 20-cone states (denoted as $20_1-$)
when $\Gamma \gtrsim 0.25$(details in Sec. VII A of the Supplemental Material).
Since the GKSL is not stable when $\Gamma \gtrsim 0.25$, we search for all stable QSL states
in the vicinity of transition point in large systems (On a $144 \times 144 \times 2$ lattice, where $144 \times 144$ corresponds to the momentum resolution, and the factor of $2$ accounts for the A and B sublattice points).
Note that there are two kinds of PSG setting in MF parameters that give rise to full symmetries in projected spin state. These
are the homogeneous setting (all parameters are position independent) and $\pi$-flux setting which originated from the vortex-free and full-vortex state in the Kitaev's exact solution.

For homogeneous setting and with $\Gamma=0.3$, the stable QSL states are given by
\begin{eqnarray} \label{homoSL} \notag
 Z_2&:&2\mbox{-},\: 8\mbox{-},\: 14_{1\mbox{-}2}\mbox{-},\: 20_{1\mbox{-}4}\mbox{-},\: 26\mbox{-} \mbox{cone}, \\ \notag
SU(2)&:&16\mbox{-},\: 32\mbox{-}  \:  \mbox{cone}, \notag
\end{eqnarray}
and for $\pi$-flux setting, the stable QSL states are given by
\begin{eqnarray} \label{piSL} \notag
&& Z_2: \pi\mbox{-}4_{1\mbox{-}2}\mbox{-}, \:, \pi\mbox{-}40\mbox{-},\:  \pi\mbox{-}56\mbox{-} \mbox{cone}, \\ \notag
&& \pi\mbox{-}28\mbox{-cone}\:\mbox{with}\: 2 \:\mbox{fermi rings}, \notag
\end{eqnarray}
where $Z_2$ and $SU(2)$ refers to the invariant gauge group (IGG)\cite{PSG_Wen} for the solution, $2,8,14...$ denote number of cones in the first
Brillouin zone(FBZ), $\pi\mbox{-}4,40,...$ denote number of cones in the folded reduced Brillouin zone due to $\pi$-flux setting,
and the subscript labels the different type QSL states with the same number of cones. The IGG determines weak gapless gauge fluctuations of the solution\cite{PSG_Wen},
where the calculation of IGG of each state is shown in Sec. V B of the Supplemental Material\cite{SMI}.

By increasing $\Gamma$, it is shown that $8$-cone state are not stable when $\Gamma$ is greater than $0.6$,
$20_4$-cone state are not stable when $\Gamma$ is greater than $0.53$,
$14_1,26$-cone states are not stable when $\Gamma$ is greater than $0.43$,
and $20_{1,3}$-cone, $\pi$-$28$-cone with $2$ fermi ring state are not stable when $\Gamma$ is greater than $0.38$.

The other states survive for large $\Gamma$ up to $1$. We further examine the stability of
2-,$14_2$-,16-,$20_2$-,$20_4$-, 32-, $\pi$-$4_{1,2}$ states
when small values of $J$ and $\Gamma'$ are turned on ($|J|,|\Gamma'|<0.1$), and found that
they are all still stable solutions.

In Fig. 5, we show the lowest fermionic energy spectrum of some of the stable QSL states we found in the homogeneous setting.
The $\chi^{\mu\nu}_{z_l}$ of each QSL state is shown in Sec. VII B of the Supplemental Material\cite{SMI}.

\subsection*{Comparison of energy obtained by  RMFT and  Monte-Carlo method}
\noindent To verify the validity of our RMFT results, we compare our RMFT results with those obtained by using Monte-Carlo (MC) method for
projection via the following procedure: \\
For a given RMFT solution with the RMFT energy $E^{\mbox{\tiny{MC}}}$ and the corresponding RMFT Hamiltonian $H^{\mbox{\tiny{RMT}}}$ in Eq.(\ref{HRMFT}), we map
$H^{\mbox{\tiny{RMF}}}$ back to ordinary fermion basis, and obtain the corresponding ground state $|\psi_0 \rangle$.
Then we use method of Monte-Carlo to treat the projection $P_G$ and compute $E^{\mbox{\tiny{MC}}}=
\frac{\langle \psi_0 | P_GHP_G|  \psi_0  \rangle }{\langle \psi_0 |P_G|  \psi_0 \rangle}$.
Similarly, we can also calculate the Wilson loop value of $\psi_0$ by using the MC method as
$W_{\mbox{\tiny{MC}}} = \frac{\langle \psi_0 | P_G \hat{W} P_G|  \psi_0  \rangle }{\langle \psi_0 |P_G|  \psi_0 \rangle}$,
where $\hat{W}$ is the product of six spin operators through hexagon edges\cite{Kitaev1}.

The ratio $E^{\mbox{\tiny{MC}}}/E^{\mbox{\tiny{RMF}}}$ and $W_{\mbox{\tiny{MC}}}$ of GKSL with different
$\Gamma$'s for  lattice size  $7\times7\times2$ are presented in TABLE I.
The error is about 1\% which shows that RMFT works excellently with high precision. In  TABLE II, we show
the ratio $E^{{\mbox{\tiny{MC}}}}/E^{{\mbox{\tiny{RMF}}}}$ and $W_{\mbox{\tiny{MC}}}$ of different QSL states with homogeneous setting at $\Gamma=0.3$.
Finally, in Fig. 6 (a), we show how energies of QSL states change versus $\Gamma$ by plotting $E^{\mbox{\tiny{RMF}}}$ relative
to energy of 2-cone state versus $\Gamma$ for different QSL states. The comparison of $E^{\mbox{\tiny{MC}}}$ and $E^{\mbox{\tiny{RMF}}}$ of different QSL states with $\Gamma/|K|=0.3$ is shown in Fig. 6 (b). It is clear that for either the RMFT energies
or the MC energies, they differs only by $\sim 0.02|K|$. In Fig. 6 (c), the phase diagram for $\Gamma'=-0.08$ is shown to
illustrate the phase transitions between the zigzag ordered state, 2-cone spin liquid state, and the polarized state.

\subsection*{Symmetry dictated  QSL states with homogeneous setting}
\noindent The quantum spin liquid phases obtained may contain Dirac cones at various $\vec{k}$ points. Here
we summarize the analysis of positions and shapes for Dirac cones from symmetry point of view. We first note that
since the Majorana fermions mix particle and hole equally, any Hamiltonian composed by quadratic
Majorana fermion terms possesses the particle-hole symmetry (PHS).
Hence energy eigenstates for particles at $-\vec{k}$ are exactly the same as the energy eigenstates for anti-particles at $\vec{k}$ with the eigen-energies satisfying $E_{-\vec{k}}=-E_{\vec{k}}$.
Therefore, in the following, we shall focus on properties at $\vec{k}$ with the understanding that
the same properties hold at $-\vec{k}$ as well.

In Fig. 7 (a), we show the momentum $\vec{k}_D$ at which the typical found Dirac cones are located for QSL states
with  homogeneous setting. It is clear that these locations are either at the high symmetry points such as $\vec{k}_K$ and $\vec{k}_{\Gamma}$ or on high symmetry axes such as $\vec{k}^{*}_{M;n}$and $\vec{k}^{**}_{M;n}$ with $n=1,2,3$. Furthermore, by using the full PSG transformation, it can be proved that the QSL states characterized by
Eq.(\ref{chizl}) must have zero energy excitation at $\vec{k}_K$(see sec. VIII B of the Supplemental Material in details\cite{SMI}). In addition, we find that
when $\chi^{0b}_{z_l}=0$, the zero energy eigenstates are formed by $b^0$;
while if $\chi^{0b}_{z_l}\neq0$, the zero energy eigenstates are formed by the
$b^0$ and $b^c \equiv \vec{b}\cdot\hat{c}=(b^x+b^y+b^z)/\sqrt{3}$. This reflects the change of the power law of field versus gap when $\hat{h}\parallel \hat{c}$.

For the Dirac cones at $\vec{k}_K$ and $\vec{k}_{\Gamma}$, since they possess the $C^{\hat{c}}_3$  rotational symmetry with respect to $c$-axis and the mirror symmetry $M^{\hat{b}}$ with respect to the $\hat{a}$-$\hat{c}$ plane,
the low energy effective Hamiltonian is given by $v_F(q_a\tau^a+q_b\tau^b)$,
where $\tau^{a,b,c}$ are the $x,y,z$ components of Pauli-matrices formed by two low energy effective states, and $\vec{q}=\vec{k}-\vec{k}_{D}$.
On the other hand, for the Dirac cones at $\vec{k}^{*}_{M;1}$ or $\vec{k}^{**}_{M;1}$,
the $C^{\hat{c}}_3$ symmetry is broken but the $C^{\hat{b}}_2$ symmetry is preserved,
hence the low energy effective Hamiltonian is given by $v^a_Fq_a\tau^a+v^b_Fq_b\tau^b$ with $v^a_F \neq v^b_F$.
Note that the global $C^{\hat{c}}_3$ symmetry is still preserved. In this case, $\vec{k}^{*}_{M;n}$ with $n=1,2,3$ are included as a whole so that under the $C^{\hat{c}}_3$ rotation, $\vec{k}^{*}_{M;1}$, $\vec{k}^{*}_{M;2}$, and $\vec{k}^{*}_{M;3}$ are cyclically permuted. The low energy Hamiltonians of the Dirac cones at $\vec{k}^{*}_{M;2,3}$ are thus generated by $C^{\hat{c}}_3$ rotations of the effective Hamiltonian at $\vec{k}^{*}_{M;1}$. For the Dirac cones at other low symmetry points, all $q_a\tau^{a}$, $q_a\tau^{b}$, $q_b\tau^{a}$, and $q_b\tau^{b}$ terms are allowed. For more details, see sec. VIII B of the Supplemental Material\cite{SMI}.

\subsection*{Symmetry dictated excitation gap-field relations}
\noindent In this subsection, we summarize the analysis of the shift in the location $\vec{k}_D$ of the Dirac cones and
the behaviour of fermionic excitation gap under weak magnetic fields $\vec{h}$. For this purpose,
we shall denote $h^a=\vec{h}\cdot\hat{a}$, $h^b=\vec{h}\cdot\hat{b}$, $h^c=\vec{h}\cdot\hat{c}$,
and $\Delta_M(\vec{k})$ as the fermionic excitation gap at $\vec{k}$.
For more details, see sec. VIII C of the Supplemental Material\cite{SMI}.

\subsubsection*{Dirac cones located at $\vec{k}_K$ point}
\noindent For $\vec{h} \parallel \hat{c}$, the magnetic field does not cause any shift $\vec{k}_D = \vec{k}_K$ and simply opens
a fermionic gap $\Delta_M(\vec{k}_K)$. We find that if $\chi^{0b}_{z_l}=0$, $\Delta_M(\vec{k}_K) \propto (h^c)^3$, while if $\chi^{0b}_{z_l}\neq0$, $\Delta_M(\vec{k}_K) \propto h^c$.

For $\vec{h} \parallel \hat{b}$,  $\vec{k}_D$ gets shifted along $\hat{a}$ axis by the fields such that
$\vec{k}'_D \propto \vec{k}_D+c_1(h^b)^2\hat{a}$, where $c_1$ is a constant. However,
the fermionic gap is not opened until two Dirac cones merge.

For $\vec{h} \parallel \hat{a}$,  $\vec{k}_D$ also gets shifted along $\hat{a}$ axis such that
$\vec{k}'_D \propto \vec{k}_D-c_1(h^a)^2\hat{a}$ with $c_1$ being the same coefficient as in $\vec{h} \parallel \hat{b}$ case.
We found that the fermionic gap follows $\Delta_M(\vec{k}'_D)\propto (h^a)^3$.
Note that for the applied weak fields along $\hat{a}$, in solving self-consistent solution in RMFT, we find that
additional terms along $c$-axis such as $g^{c}_{m,i}$, and $m^{c}$ will be generally induced.

By combining results obtained for $\vec{h} \parallel \hat{a}$ and $\vec{h} \parallel \hat{b}$ and using $C^{\hat{c}}_3$ symmetry, we find that for $\vec{h}=h(\cos\varphi\hat{a}+\sin\varphi\hat{b})$,
$\Delta_M(\vec{k}'_D) \propto h^3$, and when
$\varphi=\pi/2$, $\Delta_M(\vec{k}'_D)=0$. Thus, the azimuthal angle-dependent gap
$\Delta_M(\varphi)$ must possess
six-fold symmetry with the nodal line along $\hat{b}$.

When both $h^a$ and $h^c$ exist, we find that $\Delta_M(\vec{k}'_D)$ are composed by
$(h^a)^3$, $(h^a)^2h^c$, $(h^c)^2h^a$, and $(h^c)^3$.
Note that $(h^c)^2h^a$  is forbidden by symmetry analysis at $\vec{k}_K$, but it can be generated
by the shifted of Dirac momentum $\vec{k}'_D-\vec{k}_D$$\propto h^ah^c$, in which $\vec{k}'_D$ breaks the $C^{\hat{c}}_3$ symmetry.

\subsubsection*{Dirac cones located at $\vec{k}^{*}_{M;1,2,3}$ or $\vec{k}^{**}_{M;1,2,3}$}
\noindent Since at $\vec{k}^{*}_{M;1}$ or $\vec{k}^{**}_{M;1}$, the  $C^{\hat{c}}_3$ symmetry is broken
but the $C^{\hat{b}}_2$ symmetry is preserved,  there are more $h$ term-dependence allowed
for the gap $\Delta_M(\vec{k}'_D)$. Here we shall focus on $\vec{k}_D=\vec{k}^{*}_{M;1}$ and find:

For $\vec{h} \parallel \hat{c}$,  the location of the Dirac cone is shifted along $\hat{a}$ axis with
$\vec{k}'_D-\vec{k}_D$$\propto (h^c)^2$ and the electronic excitation gap follows
as $\Delta_M(\vec{k}'_D)\propto h^c$.

For $\vec{h} \parallel \hat{b}$,  the location of the Dirac cone is shifted along $\hat{a}$ axis with
 $\vec{k}'_D-\vec{k}_D$$\propto (h^b)^2$.  However,
the fermionic gap is not opened until two Dirac cones merge.

For  $\vec{h} \parallel \hat{a}$, the location of the Dirac cone is shifted along $\hat{a}$ axis with
$\vec{k}'_D-\vec{k}_D$$\propto (h^a)^2$.
We found that the fermionic gap follows $\Delta_M(\vec{k}'_D)\propto h^a$.

For $\vec{h}=h(\cos\varphi\hat{a}+\sin\varphi\hat{b})$, we find that
$\Delta_M(\vec{k}'_D) \propto h$ and when
$\varphi=\pi/2$, $\Delta_M(\vec{k}'_D)=0$. Furthermore,
due to the broken $C^{\hat{c}}_3$ symmetry,
the azimuthal angle-dependent gap $\Delta_M(\varphi)$ possess
two-fold reflection symmetry through $\hat{b}$ with the nodal line being along $\hat{b}$. Note that
the global $C^{\hat{c}}_3$ symmetry is restored if
we treat the gap at $\vec{k}^{*}_{M;2,3}$ as a whole.
These three gaps give rise to the six-fold symmetry versus the azimuthal angle.

\subsubsection*{Dirac cones at low symmetry points}
\noindent Generally, $\Delta_M(\vec{k}'_D)$ is proportional to $h$. For individual Dirac cone at
low symmetry point, there is no special symmetry pattern in the corresponding azimuthal angle-dependent
gap $\Delta_M(\varphi)$. However, if we include all Dirac cones at all $C^{\hat{c}}_3$ and $M$ symmetry partner
points of $\vec{k}'_D$, the symmetry will restore.

\subsubsection*{Verification of symmetry dictated gap-field relations in RMFT}
\noindent To illustrate the symmetry dictated gap-field relations in RMFT, we first choose the $14_2$-cone state and examine
the angle-dependent gap of each Dirac cone under weak field. As shown in Fig. 7 (b)-(e),  when all  $C^{\hat{c}}_3$ and $M$ symmetry partner
points to $\vec{k}'_D$ are included,  the fermionic gaps $\Delta_M(\vec{k}'_D)$ exhibit the six-fold symmetry pattern, in agreement with the analysis.

To check the gap-field relation, we consider the $8$-cone state and the $2$-cone state, in  which
$\chi^{0b}_{z_l}\neq0$ for $8$-cone state and $\chi^{0b}_{z_l}=0$ for $2$-cone state.
As shown in Fig. 8 (a), for $\vec{h} \parallel \hat{c}$, we find that $\Delta_M(\vec{k}'_K)$ from the $2$-cone state is proportional to $(h^c)^3$, while for other cone state,  $\Delta_M(\vec{k}'_K)$ is proportional to $h^c$.
On the other hand, as shown in Fig. 8 (b), for $\vec{h} \parallel \hat{a}$, we find that the fermionic gap for both states
are proportional to $(h^a)^3$, for other cone-states, $\Delta_M(\vec{k}'_K)$ is proportional to $h^a$. These and all other RMFT results are
consistent with the symmetry analysis.

\subsection*{The observed QSL state under magnetic fields in the $\hat{a}$-$\hat{b}$ plane}
\noindent To explain the experimental observation on the QSL state in the candidate Kitaev material
$\alpha$-$\mbox{RuCl}_3$\cite{Matsuda_Nat,Matsuda_Sci,Matsuda_PRB}, we investigate the band Chern number and fermionic excitation gap for different QSL states under general magnetic fields $\vec{h}$.

We first note that, for a given magnetic field, different fermionic gaps may open at different $\vec{k}$ points, but the low-temperature behavior of the specific heat is governed by the smallest gap.  For example, from symmetry analysis on a single cone,  only the gap of cone at $\vec{k}_K$ is proportional to $(h^a)^3$, all other cones'
gaps is proportional to $h^a$. The experimentally observed gap, however, is well fitted by $(h^a)^3$\cite{Tanaka}.
Hence it may seem that multi-cone states are excluded. However, as explained, the low temperature behaviour of specific heat is dominated by the minimum gap in FBZ, hence if the gap near $\vec{k}_K$ is much smaller than gaps of other cones,
the specific heat that is proportional to $(h^a)^3$ is still possible. Since in the real material, the strength of Kitaev's interaction is estimated
in $100\sim200$$K$\cite{KStrength_1,KStrength_2}, the half-quantized thermal Hall phase occurs around $h/|K| \approx 0.05\sim0.1$\cite{Matsuda_Sci,WeiLi}.
We examine  gaps of cones from multi-cone QSL states, and find that around this field range the different cones' gap are in the same
order. Hence, multi-cone states are generally excluded(except for $\pi\mbox{-}4_{1,2}$-cone state).

Secondly, we find that when the applied field is along $\hat{a}$ direction, only the 2-cone state has non-trivial topology with  Chern number $\nu=1$. At $\Gamma=0.3$, the $\nu=1$ phase of the 2-cone state is present in the range $h/|K|\lesssim 0.1$ and for $h/|K| > 0.1$, the system becomes a polarized state.
We have also checked these properties under small $|J|$ and $|\Gamma'|$ ($<0.1$), where only minor quantitative differences are observed.
For the real material, the upper boundary of half-quantized thermal Hall phase is also around $h/|K| \sim 0.1$\cite{Suetsugu}.
Furthermore, as shown in Fig. 9 (a), the field-azimuthal-angle dependence of the fermionic gap for the $2$-cone state also shows good agreement with the experimental observation\cite{Tanaka}.
While for other QSL states, the field-azimuthal-angle dependence does not show simple patterns.
Therefore we conclude that the system should be in this 2-cone state we found.
We further characterize the found 2-cone state by examining its ground state degeneracy (GSD) on a torus.
The GSD on a torus refers to the fact that inserting a global $Z_2$ $\pi$-flux into either hole of a torus costs no energy in the thermodynamic limit\cite{TP_Wen}.
This procedure is equivalent to changing the boundary condition of the mean-field Hamiltonian from periodic to anti-periodic. Accordingly,
we can label the corresponding many-body states as $|\psi_{\pm\pm}\rangle$, where the subscripts denote the boundary conditions along the x- and y-directions, respectively.
After projecting these four states into the physical spin Hilbert space, the number of linearly independent states determines to the GSD on a torus.
In practical computations, we first transform the Majorana fermion mean-field Hamiltonian into an ordinary fermionic basis. We then obtain the Gutzwiller-projected many-body spin states corresponding to the four boundary conditions on the $8\times8\times2$ lattice and compute the GSD. For the 2-cone state under a weak magnetic field, the GSD on a torus is found to be 3, since $|\psi_{++}\rangle$ has odd fermionic parity and thus vanishes after Gutzwiller projection. This result indicates that the 2-cone state is a topologically ordered quantum spin liquid with long-range entanglement.\cite{TP_Wen,Wang_PRL}.

Because only GKSL and the found 2-cone state hold Chern number $\nu=\pm 1$ and both states
show simple patterns of the field-azimuthal-angle dependence of the gap, we will focus on further comparisons of these two states.
As shown in Fig. 9 (e), the found 2-cone state has different lowest spinon spectrum
from that of the GKSL 2-cone state analytically continued from the original Kitaev solution shown in Fig. 9 (d).
Furthermore, as shown in Fig. 9 (b) and (c), while the found 2-cone state gives the same angle-dependent energy gap observed by experiments, the Chern number pattern for six sectors of angles is opposite in sign in comparison to those of the original Kitaev spin liquid. In addition, the value of the Wilson loop for the found 2-cone state with $\Gamma =0.3$ is -0.353 in contrast to the positive values of the Wilson loop of GKLS state (see Table II).

Experimentally, the sign of the Chern number determines the sign of the
thermal Hall conductivity $\kappa_{xy}$. Hence when the applied field is along $\hat{a}$, while our theory predicts a negative $\kappa_{xy}$ for GKSL, it predicts a positive $\kappa_{xy}$ for the found $2$-cone state. This difference should serve as a check point in experiments. Nonetheless, the determination of sign for experimentally observed
$\kappa_{xy}$ is ambiguous. In particular, when the applied field is along $\hat{a}$, some experiments reported negative $\kappa_{xy}$\cite{Matsuda_Sci},
while some experiments reported positive $\kappa_{xy}$\cite{Bruin,Czajka_2}. The difference may result from that either the applied field is not in the correctly aligned in the $+\hat{a}$ direction or when extracting the thermal Hall conductivity from experimental data, some additional minus sign might be included.  Hence the correct sign of the Chern number is an issue to be clarified by future experiments.

Finally, we investigate how gaps close nearby the polarized phase boundary for these two states.
As shown in Fig. 10, we find that the fermionic gap for the GKSL state closes at $\vec{k}_{\Gamma}$, while as shown in Fig. 11, the fermionic gap for the found $2$-cone state closes at $\vec{k}_{M}$ and $\vec{k}_{Y}$. At the fields beyond those plotted in Fig. 10 and Fig. 11, the states are topologically
trivial, $\nu=0$ and with no long range entanglements, so they are just trivial states, similar to the spin-polarized state.
Clearly, we find that  the critical field for the GKSL state is much smaller than that for the 2-cone state, which can be attributed the fact that the GKSL state is unstable when $\Gamma \gtrsim 0.25$. Therefore, being different from the original Kitaev spin liquid, the 2-cone state predicted by our RMFT theory is a new state emerging for Kitaev materials in external magnetic fields for large $\Gamma$.

The above analysis suggests that instead of the GKLS state being the state observed in experiments, the found 2-cone state should be the state observed by experiments. To further check and confirm this conclusion, we propose to check
the field-direction-dependence of the fermionic excitation gap in polar angles. In Fig. 12 (a), we compare the polar-angle $\theta$ dependence of fermionic gap
for the 2-cone state and the GKLS state. Clearly, the polar-angle dependence for the 2-cone state is more symmetric,
while the asymmetric feature of GKSL state is originated from the KSL state. In Fig. 12 (b), we show how the polar-angle dependence for
the 2-cone state change for large field and different $\Gamma$s, which can serve as an important check in experiments.

\section*{Discussion}
\noindent Although many experimentally observed  features such as  sign-changing of thermal Hall conductivity
and the fitted band gap can be naturally explained in terms of quantum spin liquids formed by Majorana
fermions\cite{Tanaka,Imamura}, the key feature, the half-integer thermal Hall conductivity, was only observed in certain magnetic-field
and temperature regimes. In particular, excessive thermal Hall conductivity was observed in high temperatures, while the thermal Hall conductivity is suppressed in low temperature regime. Theoretically, while excess thermal Hall conductivity  could be attributed to additional
channels of contribution from phonons or magnons, the suppression of  the thermal Hall conductivity in low temperatures requires an explanation\cite{Czajka_1,Bruin,Lefran,Czajka_2,Kee1}.
The phonon-edge channel decoupling mechanism was proposed to explain the non-quantized thermal Hall conductivity at very low temperature\cite{PhononHall1,PhononHall2}.
In this scenario, the measured thermal Hall conductivity can be either larger or smaller than the half-quantized value, depending on the thermal contact properties.
However, this mechanism does not provide a quantitative description that can be directly compared with experimental data. In contrast, both experimental groups reported a suppression of the thermal Hall conductivity in the $3\sim10$$K$ temperature range\cite{Czajka_1,Bruin}.While the decoupling effect might emerge below 3$K$, as the measurements become relatively unreliable in this regime. In the fermion-interpretation scenario, the suppression could be due to the opening of a fermionic gap in the edge states. Here we provide an explanation to quantitatively describe the observed thermal Hall conductivity in the $3\sim 10$$K$ temperature range.
Indeed, the thermal Hall current  is given by $I_E = \int_{\epsilon(q)\geq 0} n(q) \epsilon(q) v(q) dq/ 2 \pi$\cite{Kitaev1}, where
$n(q) = 1/(1+e^{\epsilon (q)/k_BT})$ is the Fermi-Dirac distribution function and $v(q)=\frac{d\epsilon (q)}{dq}$ with $\epsilon(q)$ being  the energy of the quasi-particle in the edge state.
If the energy dispersion of the edge state opens a gap $\Delta$ and the edge state is cutoff by the bulk state at $\epsilon=\Lambda$,
the finite temperature thermal Hall current is given by
\begin{eqnarray} \label{thermal_current}
I_E = \frac{k^2_B}{2\pi\hbar} \int^{\Lambda/k_BT}_{\Delta/k_BT} \frac{x}{1+e^{x} } d x.
\end{eqnarray}
After performing the integration and using $\kappa^{2D}_{xy}=\frac{dI_E}{dT}$, we find that the thermal Hall conductance is given by
\begin{eqnarray} \label{kappa}
\kappa^{2D}_{xy} = T\left(\frac{\pi}{6}\frac{k^2_B}{\hbar}\right)\left( F(\frac{\Lambda}{k_BT}) -  F(\frac{\Delta}{k_BT})\right),
\end{eqnarray}
where $F(x)=\frac{6}{\pi^2} \left[ Li_2(-e^{-x}))-x \ln(1+e^{-x})- \frac{1}{2}\frac{x^2}{1+e^{x}} \right]$. As shown in Fig.~13 (a),(b) experimental
$\kappa^{exp}_{xy}/T$ data from two different groups\cite{Bruin,Czajka_2} both can be well fitted by using Eq.(\ref{kappa}) plus $T^2$ contribution from phonons\cite{Lefran,PhononHall2,PhononHall3,PhononHall4,PhononHall5}, even though the fitted gap is a bit large and the fitted phonon contribution could also be overestimated
in current theory suggestion.
We speculate therefore that, due to enhanced density of Majorana fermions and additional interactions not included in the present paper
such as interlayer coupling, edge states become gapped with thermal Hall conductivity suppressed at low temperatures, while the bulk
gap for thermodynamics remains the one described by the model described in the main text. Note that the fitted $\Lambda$ values exhibit non-monotonic behavior versus
magnetic fields, which can understood by examining how the minimum gap at $k_K$, $k_M$ and $k_Y$ points vary versus magnetic fields in Fig. 11 and Supplementary Fig. 14.
Since $\Lambda$ is determined by how the edge state merges into the bulk bands and are determ,ined by the bulk gap. The non-monotonic behavior originates
from the non-monotonic behavior of the bulk gap determined by the minimum gap at $k_K$, $k_M$ and $k_Y$ of the found 2-cone state.

In conclusion, we develop a gauge-invariant renormalized mean-field theory that is applicable to realistic Kitaev materials and yields reliable phase diagram in thermodynamic limit. In particular, while our RMFT reproduces previous results based on numerical methods, it also predicts several new stable QSL states with energies being very close and differing by about $0.01\sim 0.02 |K|$.  Among those stable QSL states, we find that a new exotic 2-cone state can account for the major experimental observations. This 2-cone state holds negative values for the Wilson loop and in the presence of magnetic fields in the $\hat{a}$-$\hat{b}$ plane, it gives rise to the same field-angle-dependent energy gap observed by experiments. However, the Chern number for six sectors of angles that it holds has opposite sign pattern in comparison to those of the original Kitaev spin liquid.  The found 2-cone state also displays non-Abelian statistics in weak magnetic fields when gap is opened and thus appears to be the state observed by experiments. We further explore more connections of our results with experiments and speculate that due to enhanced density of Majorana fermions and additional interactions not included, such as interlayer coupling, edge states become gapped so that the thermal Hall conductivity gets suppressed at low temperatures, as observed in experiments. In addition, we show that the polar-angle dependence of fermionic gap can be used to distinguish the found 2-cone state from the  KSL state.


\section*{Methods}
\noindent As mentioned in the introduction, the method of the renormalized mean-field theory was a reliable method introduced to find
correlated electronic phases in the Hubbard model for high-Tc cuprate superconductors. The theory has achieved great success for systems being spin-SU(2) invariant. Here we shall
further generalize RMFT to systems without spin-SU(2) symmetry by combining RMFT with the gauge-invariant decomposition of spin, and derive a gauge-invariant RMFT
that goes beyond the usual approach based on the Gutzwiller approximation, using local classical weight factors\cite{RMFT1}.
The idea is to replace  classical weights by quantum averages over local clusters of operators. More precisely, the quantum average is performed
in the unprojected Fock space so that the effects of the projection are included in the effective Hamiltonian
in unprojected Fock space through renormalized factors.  If $ |\psi_0  \rangle$ is in the class of wavefunctions
in the unprojected Fock space that we look for optimizing the ground state energy, the average of the Hamiltonian $\frac{\langle \psi_0 | P_GHP_G|  \psi_0  \rangle }{\langle \psi_0 |P_G|  \psi_0 \rangle}$ in the projected Fock space, is approximated by
\begin{eqnarray} \label{Ev1}
E^{\mbox{\tiny{RMF}}}_v = \sum_{\langle ij \rangle, \alpha\alpha^{\prime}} g^{\alpha\alpha^{\prime}}_{J,ij} J^{\alpha\alpha^{\prime}}_{ij}
\langle \sigma^{\alpha}_{i} \sigma^{\alpha^{\prime}}_{j} \rangle_0 -\sum_{i\alpha}\frac{h^{\alpha}}{2}g^{\alpha}_{m,i}m^{\alpha}_i.
\end{eqnarray}
Here $g_O's$ are renormalized factors defined as the ratio of  the average of the corresponding operator $O$ in projected Fock space to the average in unprojected Fock space as  $\langle O \rangle / \langle O \rangle_0$, where the average in $ \langle O \rangle_0$ is the average over unprojected state $|\psi_0 \rangle$, while the average in $ \langle O \rangle$ is the average over projected state $P_G|  \psi_0  \rangle$ with $P_G$ being the Gutzwiller projector given by $P_G=\prod_{j=1}^{N_s} \frac{1+D_j}{2}$ with $N_s$ being number of sites and $D_j=b_j^xb_j^yb_j^zb_j^0 = (2 n_{j \uparrow}-1)(1-2n_{j \downarrow})$\cite{KitaevPhy1,KitaevPhy2}. The self-consistent RMFT Hamiltonian $H^{\mbox{\tiny{RMF}}}$ is obtained by the minimization of $E^{\mbox{\tiny{RMF}}}_v$\cite{RMFT7} with constraint $\langle Q^{\alpha}_i \rangle=0$\cite{PSG_Wen}, given by
\begin{eqnarray} \label{HRMFT}
H^{\mbox{\tiny{RMF}}}=\sum_{ij,\mu v}\frac{\partial E_v}{\partial \chi_{ij}^{\mu v} }\hat{\chi}_{ij}^{\mu v}+\sum_{i,\alpha}\frac{\partial E_v}{\partial m_{i}^{\alpha} } \sigma_i^{\alpha}
+\sum_{i,\alpha}\lambda^{\alpha}_{i,Q}Q_i^{\alpha},
\end{eqnarray}
where $\chi_{ij}^{\mu v}=\langle \hat{\chi}_{ij}^{\mu v} \rangle_0$, $\hat{\chi}_{ij}^{\mu v}=ib^{\mu}_ib^{v}_j$,
$m_i^{\alpha}=\langle  \sigma_i^{\alpha}\rangle_0$,  and $\lambda^{\alpha}_{i,Q}$ are the  Lagrangian multipliers
that enforce $\langle  Q_i^{\alpha} \rangle_0=0$.
Note that taking $| \psi_0 \rangle$ as the ground-state wavefunction of the mean-field Hamiltonian in the absence of magnetization, it generally has the property that averages of unpaired Majorana fermion $\langle b_i^{\alpha} \rangle_0$ and pairing on the same site $ \langle  b^{\mu}_{i} b^{\nu}_{i} \rangle_0$ vanish.

To find the renormalized $g_O$ factors, we note that using the identity  $D_j\sigma_j^{\alpha}=\sigma_j^{\alpha}D_j$,
we have $P_G \sigma_i^{\alpha}\sigma_j^{\alpha^{\prime}} P_G$ = $\sigma_i^{\alpha}\sigma_j^{\alpha^{\prime}} P_G$, and
thus $g^{\alpha\alpha^{\prime}}_{J,ij}$ and $g^{\alpha}_{m,i}$ factors can be expressed as
\begin{eqnarray} \label{gDef}
g^{\alpha\alpha^{\prime}}_{J,ij}=\frac{\langle \sigma_i^{\alpha}\sigma_j^{\alpha^{\prime}}  P_G\rangle_0 }
{\langle P_G\rangle_0 \langle \sigma_i^{\alpha}\sigma_j^{\alpha'} \rangle_0 },
g^{\alpha}_{m,i}=\frac{\langle \sigma_i^{\alpha}P_G\rangle_0 }
{\langle P_G \rangle_0 m^{\alpha}_i },
\end{eqnarray}
where $m^{\alpha}_i=\langle \sigma_i^{\alpha} \rangle_0$. We note in passing that in general, the renormalized $g_O$ factors are combinations of more fundamental renormalization of the field $\sigma^{\alpha}$ and the spin-spin interaction $J$.

As an illustration, we will first evaluate $g^{\alpha\alpha^{\prime}}_{J,ij}$ in the Kitaev model.  First, by using the Wick's theorem, the average of products of $D_j$  on any finite subset of the lattice,  $\langle \psi_0 | \prod_{j_1,j_2,...} D_{j_1}D_{j_2}... | \psi_0  \rangle $, must contain  $ \langle b^{\mu}_{i} b^{\nu}_{i}  \rangle_0$
, $\langle b_i^{\alpha} \rangle_0$, $ \langle b^{0}_{i} b^{\alpha}_{j} \rangle_0$, $ \langle b^{\alpha}_{i} b^{\alpha'}_{j}  \rangle_0$($\alpha \neq \alpha'$),  and thus they must vanish. Only the average $\langle \psi_0 | D | \psi_0 \rangle$ survives and is equal to $\langle \psi_0 | \prod^{N_s \rightarrow \infty}_{j}D_j| \psi_0  \rangle =1$\cite{KitaevPhy2}. This results from $D$ is the total fermion parity operators, and thus commutes with Kitaev model under Kitaev's decomposition,
gives $D =\pm 1$ for eigenstates and then gives $D=1$ for physical spin ground state.
Hence $\langle P_G \rangle_0 = (1+ \langle \psi_0 | D | \psi_0  \rangle)/(2^{N_s})=1/(2^{N_s-1})$.
To evaluate $\langle \sigma_i^{\alpha}\sigma_j^{\alpha'}  P_G  \rangle_0$, we focus on projection operators $D_i$ and $D_j$ and the projection operators on complementary lattice points $\bar{D_i}$ and $\bar{D_j}$ (complementary lattice points to lattice point $i$ is the collection of whole lattice points that removes lattice point $i$ )so that $P_G$ can be generally rewritten as
\begin{widetext}
\begin{eqnarray}
2^{N_s} P_G  &= & 1+ D_i +D_j +D_i D_j + \overline{D_i} + \overline{D_j} +\overline{D_iD_j}  + \sum_{\{ i_1,i_2, \cdots \}} \prod_{i1,i2, ...\neq i,j } D_{i_1} D_{i_2} \cdots  \nonumber \\
&=& (1+D_i)(1+D_j)(1+D)+ R_G, \nonumber \\ \label{P_G_1}
\end{eqnarray}
where $R_G \equiv \sum_{\{ i_1,i_2, \cdots \}} \prod_{i1,i2, ...\neq i,j } D_{i_1} D_{i_2} \cdots$ is the remaining products of $D_j$ operators and the second equation follows by using $DD_i=D_iD=\overline{D_i}$ and $DD_iD_j=D_iD_jD=\overline{D_iD_j}$. For
the gauge-invariant decomposition of spin, we have $\sigma_i^{\alpha} D_i=  \sigma_i^{\alpha}$. Therefore, we obtain
\end{widetext}
\begin{equation}
\langle \sigma_i^{\alpha}\sigma_j^{\alpha'}  P_G  \rangle_0 =   \frac{1}{2^{N_s}} \langle \sigma_i^{\alpha}\sigma_j^{\alpha'} (2^3+ R_G) \rangle_0. \label{spin_spin}
\end{equation}
Clearly,  by using the Wick's theorem to decompose the second term on the right-hand side into linked clusters that contain the factor $\langle \sigma_i^{\alpha}\sigma_j^{\alpha'}  \rangle_0$, the coefficient to $\langle \sigma_i^{\alpha}\sigma_j^{\alpha'}  \rangle_0$ contains averages of
products of $D_j$ on finite subset of the lattice and thus must vanish. Therefore, we obtain
\begin{equation}
\langle \sigma_i^{\alpha}\sigma_j^{\alpha'}  P_G  \rangle_0  =  \frac{1}{2^{N_s-3}} \left( \langle \sigma_i^{\alpha}\sigma_j^{\alpha'} \rangle_0 + \langle \sigma_i^{\alpha}\sigma_j^{\alpha'} R_G \rangle_c \right), \label{spin_spin_1}
\end{equation}
where $\langle \sigma_i^{\alpha}\sigma_j^{\alpha'} R_G \rangle_c$ denotes terms with projectors $D_j$ connected to $\sigma_i$ or $\sigma_j$ in the decomposition through the Wick's theorem.  For the Kitaev model, because $R_G$ excludes $\overline{D_i}$, $\overline{D_j}$ and $\overline{D_iD_j}$, products of $D_j$ that connect to  $\sigma_i$ or $\sigma_j$ only occupy on finite subset of the lattice and their averages contain  $ \langle b^{\mu}_{i} b^{\nu}_{i}  \rangle_0$ or $\langle b_i^{\alpha} \rangle_0$, and thus they must vanish.  As a result, we find that $g^{\alpha\alpha^{\prime}}_{J,ij}=4$
for the Kitaev model.
The above evaluation of $g^{\alpha\alpha^{\prime}}_{J,ij}$ for the Kitaev model can be generalized
to evaluate $g^{\alpha\alpha^{\prime}}_{J,ij}$ for the generalized Kitaev model in the presence
of magnetic field.  In this case,  $ \langle b^{\mu}_{i} b^{\nu}_{i}  \rangle_0$ is nonzero due to the presence of magnetization
but $\langle b_i^{\alpha} \rangle_0$ still vanishes. Furthermore, $\langle \sigma_i^{\alpha}\sigma_j^{\alpha'} R_G \rangle_c$
in Eq.(\ref{spin_spin_1})  no longer vanishes.  To take it into account, we include more lattice points in $P_G$ such that in the minimum inclusion, nearest neighboring points shown in Fig. 1 (b) and (c) are included. As a result, after rearranging $P_G$,  Eq.(\ref{P_G_1}) is rewritten as
\begin{widetext}
\begin{eqnarray}
&&2^{N_s} P_G  = (1+D_i)(1+D_j)(1+D_{i_1})(1+D_{i_2})(1+D_{j_1})(1+D_{j_2})(1+D)+ R_{G_1}, \nonumber \\
&& = (1+D_i)(1+D_j)(1+D_{j_1})(1+D_{j_2})(1+D_{j_3})(1+D)+ R_{G_2},
\label{P_G_2}
\end{eqnarray}
where the first equation is for the evaluation of $g^{\alpha\alpha^{\prime}}_{J,ij}$ and the second equation is for the evaluation of $g^{\alpha}_{m,i}$.
Hence by using $\sigma_i^{\alpha} D_i=  \sigma_i^{\alpha}$, $g^{\alpha\alpha^{\prime}}_{J,ij}$ and $g^{\alpha}_{m,i}$ are given by
\begin{eqnarray}
&&g^{\alpha\alpha^{\prime}}_{J,ij} = \frac{4 \langle \sigma_i^{\alpha}\sigma_j^{\alpha'}(1+D_{i_1})(1+D_{i_2})(1+D_{j_1})(1+D_{j_2})  \rangle_0}{\langle \sigma_i^{\alpha}\sigma_j^{\alpha'} \rangle_0 \langle (1+D_i)(1+D_j)(1+D_{i_1})(1+D_{i_2})(1+D_{j_1})(1+D_{j_2}) \rangle_0}, \nonumber \\
&&g^{\alpha}_{m,i} = \frac{2 \langle \sigma_i^{\alpha} (1+D_{j_1})(1+D_{j_2})(1+D_{j_3})  \rangle_0}{\langle \sigma_i^{\alpha} \rangle_0 \langle (1+D_i)(1+D_{j_1})(1+D_{j_2})(1+D_{j_3}) \rangle_0}.
\end{eqnarray}
By using the Wick's theorem,  $g^{\alpha\alpha^{\prime}}_{J,ij}$ and $g^{\alpha}_{m,i}$ are evaluated in details
in Sec. IV E of the Supplemental Material\cite{SMI} and are given by
\begin{eqnarray}
g_{m,i}^{\alpha}
=\frac{2}{m_i^{\alpha}}\times\frac{\tilde{m}_i^{\alpha}}{\tilde{\rho}_i},\:\:
g_{J,ij}^{\alpha\alpha^{\prime}}
=\frac{4}{\langle\sigma_i^{\alpha}\sigma_j^{\alpha'}\rangle_{0}} \times
\frac{\langle\sigma_i^{\alpha}\sigma_j^{\alpha'}\rangle_{\chi^2}+\tilde{m}_i^{\alpha}\tilde{m}_j^{\alpha'}-m^{dc}_{ij}}
{\langle D_i D_j \rangle_{\bar{\chi}}+\tilde{\rho}_i\tilde{\rho}_j-\rho^{dc}_{ij}},
\end{eqnarray}
where $m^{\alpha}_i$ is the bare moment, $\rho_i=1+m^2_i=\langle 1+D_i \rangle_0$ is twice of the bare local spin wave-function weight,
\begin{eqnarray}
\tilde{m}_i^{\alpha}=m_i^{\alpha}+\sum_{\langle i,j_n \rangle}\frac{\langle \sigma_i^{\alpha}D_{j_n} \rangle_{\bar{\chi}}}{\rho_{j_n}}\:\:\:\mbox{and}\:\:\:
\tilde{\rho}_i= \rho_{i}+\sum_{\langle i,j_n \rangle} \frac{\langle D_i D_{j_n} \rangle_{\bar{\chi}}}{\rho_{j_n}}
\end{eqnarray}
are the renormalized moment $\tilde{m}^{\alpha}_i$ and the renormalized local spin wave-function weight $\tilde{\rho}_i$ by including the correlation from all adjacent sites.
$\langle \sigma_i^{\alpha}\sigma_j^{\alpha^{\prime}}\rangle_0 =\langle \sigma_i^{\alpha}\sigma_j^{\alpha^{\prime}}\rangle_{\chi^2}+m_i^{\alpha}m_j^{\alpha^{\prime}}$, and $\langle \sigma_i^{\alpha}D_j \rangle_{\bar{\chi}} =2\sum_{\alpha'}m_j^{\alpha^{\prime}}\langle \sigma_i^{\alpha}\sigma_j^{\alpha^{\prime}}\rangle_{\chi^2}$  with
\begin{eqnarray} \notag
&&\langle D_i D_j \rangle_{\bar{\chi}}=4\sum_{\alpha\alpha^{\prime}}m_i^{\alpha}m_j^{\alpha^{\prime}}\langle \sigma_i^{\alpha}\sigma_j^{\alpha^{\prime}}\rangle_{\chi^2}
+\begin{vmatrix} \chi_{ij}^{00} & \chi_{ij}^{0x} & \chi_{ij}^{0y} & \chi_{ij}^{0z}\\
                                                \chi_{ij}^{x0} & \chi_{ij}^{xx} & \chi_{ij}^{xy} & \chi_{ij}^{xz}\\
                                                \chi_{ij}^{y0} & \chi_{ij}^{yx} & \chi_{ij}^{yy} & \chi_{ij}^{yz}\\
                                                \chi_{ij}^{z0} & \chi_{ij}^{zx} & \chi_{ij}^{zy} & \chi_{ij}^{zz} \end{vmatrix},\\
&\text{and }&\langle \sigma_i^{\alpha}\sigma_j^{\alpha^{\prime}}\rangle_{\chi^2} =\frac{1}{4} \left(-\chi_{ij}^{00}\chi_{ij}^{\alpha\alpha^{\prime}}+\chi_{ij}^{0\alpha'}\chi_{ij}^{\alpha0}
+\chi_{ij}^{0\gamma^{\prime}}\chi_{ij}^{\alpha\beta^{\prime}}-\chi_{ij}^{0\beta^{\prime}}\chi_{ij}^{\alpha\gamma^{\prime}}
+\chi_{ij}^{\gamma0}\chi_{ij}^{\beta\alpha^{\prime}}-\chi_{ij}^{\gamma\alpha^{\prime}}\chi_{ij}^{\beta0}
-\chi_{ij}^{\gamma\gamma^{\prime}}\chi_{ij}^{\beta\beta^{\prime}}+\chi_{ij}^{\gamma\beta^{\prime}}\chi_{ij}^{\beta\gamma^{\prime}}\right). \nonumber \\
\end{eqnarray}
\end{widetext}
The $m^{dc}_{ij}$ and $\rho^{dc}_{ij}$ denotes the double-counting and high $\chi$ order ($\chi^6,\chi^8,...$) term.
Note that the above averages $\langle \sigma_i^{\alpha}\sigma_j^{\alpha^{\prime}}\rangle_0$, $\langle \sigma_i^{\alpha}D_j \rangle_{\bar{\chi}}$, and $\langle D_i D_j \rangle_{\bar{\chi}}$ are all gauge invariant, hence $g^{\alpha\alpha^{\prime}}_{J,ij}$ and $g^{\alpha}_{m,i}$ are also gauge invariant.
In QSL phase under zero magnetic field without spontaneous magnetic ordering,
the $g_{J,ij}^{\alpha\alpha^{\prime}}$ factor becomes simpler and is independent of $\alpha\alpha^{\prime}$ as follows
\begin{widetext}
\begin{eqnarray} \label{SLgJ}
g_{J,ij}^{SL}=\frac{4}{1+\langle D_iD_j \rangle_{\bar{\chi}}+\langle D_iD_{j_1} \rangle_{\bar{\chi}}
+\langle D_iD_{j_2} \rangle_{\bar{\chi}}+\langle D_jD_{i_1} \rangle_{\bar{\chi}}+\langle D_jD_{i_2} \rangle_{\bar{\chi}}}.
\end{eqnarray}
\end{widetext}
Clearly, it reduces back to $4$ for the Kitaev model since all $\langle D_iD_j \rangle_{\bar{\chi}}$ equal to zero for Kitaev's exact ground state.

\subsection*{Data availability}
\noindent Experimental data extracted from references for Fig.~13 are available at https://github.com/ChouPoHao/ExtractedThermalHallData .
Codes are available on request from the corresponding author.
\subsection*{Acknowledgements}
\noindent This work gets support from Taiwan Consortium of Emergent Crystalline Materials (TCECM) project that is sponsored by National  Science and Technology Council (NSTC) in Taiwan. We also acknowledge support from the Center for Quantum Science and Technology (CQST) within the framework of the Higher Education Sprout Project by the Ministry of Education (MOE) in Taiwan.
\subsection*{Author contributions}
\noindent SY and CCH coordinated the project. PHC, CYM and SY conducted the computation and developed the theory.  PHC, CYM, and SY wrote the manuscript with the input from CCH.  \\

\subsection*{Competing interests}
\noindent Authors declare  no competing interests.

\begin{figure*}[htbp]
\includegraphics[height=2.5in,width=3.1in] {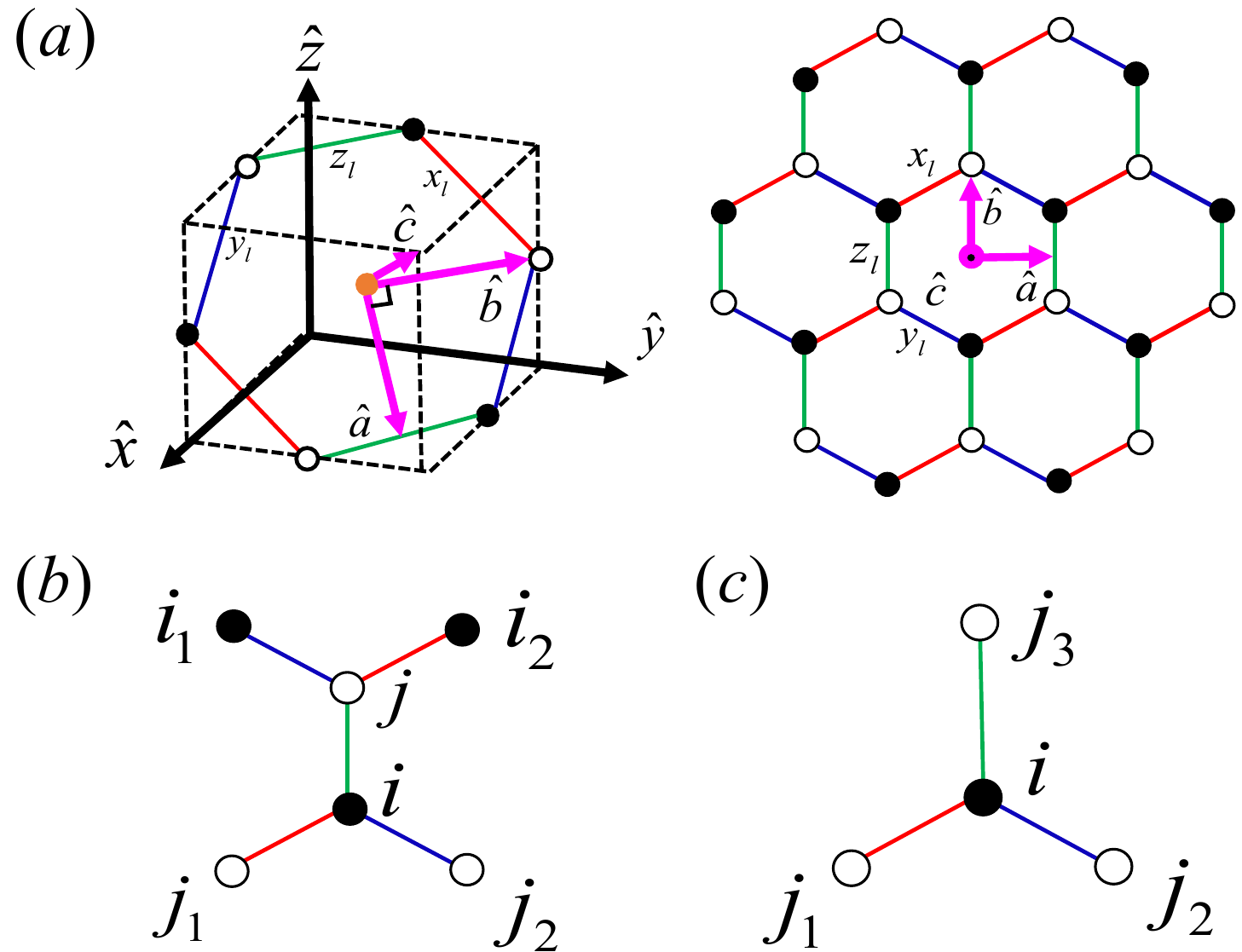}
\caption{ Honeycomb lattice and linking clusters used in RMFT: (a) The schematic honeycomb lattice with labeled links, where $\hat{a}=[11\bar{2}]$, $\hat{b}=[\bar{1}10]$, $\hat{c}=[111]$ and all are unit vectors.
(b) Linking cluster used to calculate $g^{\alpha\alpha^{\prime}}_{J,ij}$. (c) Linking cluster used to calculate $g^{\alpha}_{m,i}$.}
\end{figure*}

\begin{figure*}[htbp]
\includegraphics[height=2.6in,width=3.4in] {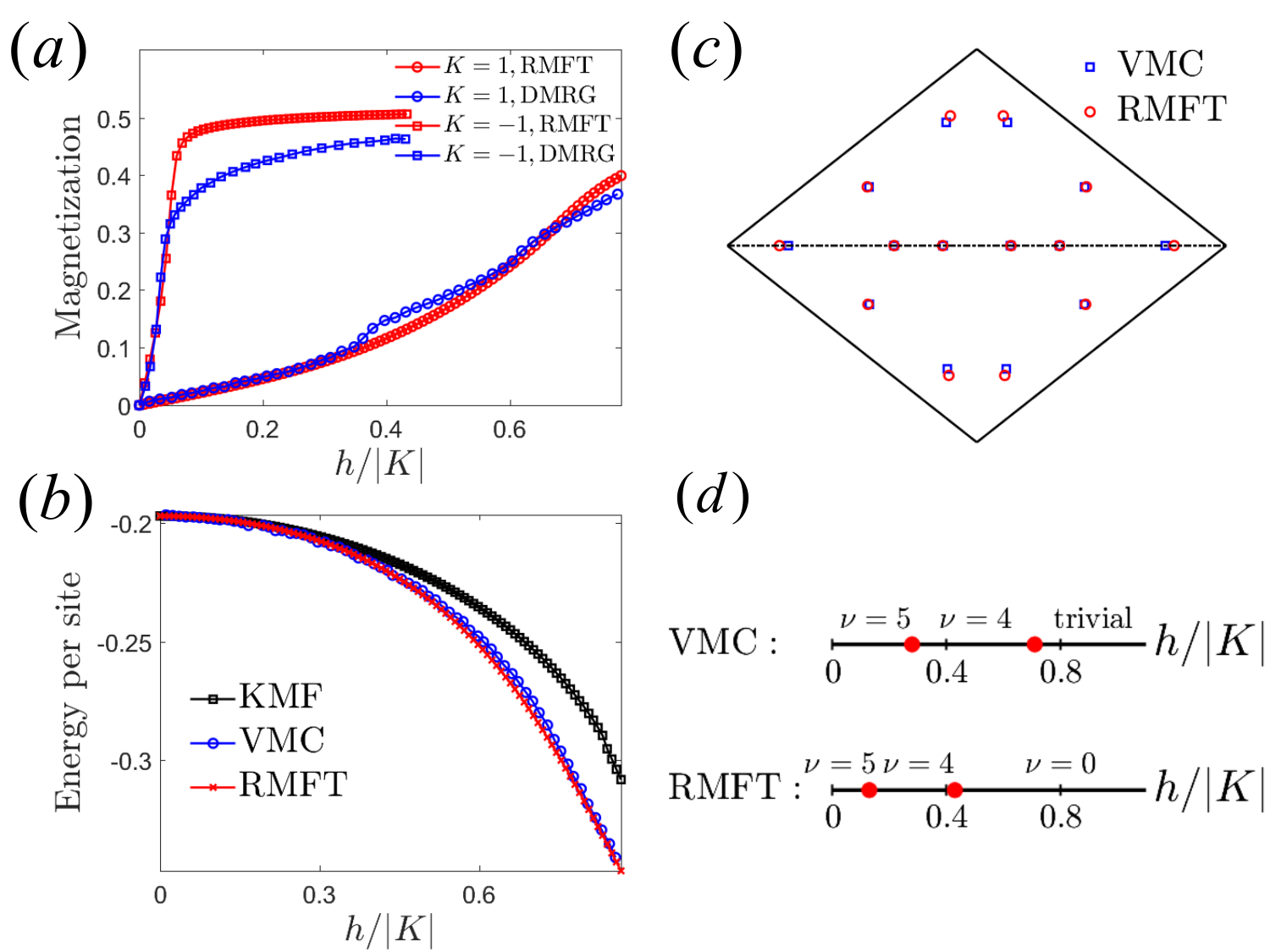}
\caption{ Comparison of RMFT with other numerical methods: (a) Comparison between RMFT and DMRG\cite{RMFT_compare} for calculated spin magnetization of the ferromagnetic or antiferromagnetic  Kitaev model under field $\vec{h}=h\hat{c}$, where $\hat{c}$ is the unit vector along $[111]$ direction. (b) Energy comparison between RMFT, KMF and VMC\cite{Jiang_PRL} of the $K>0$ (antiferromagnetic) Kitaev model under field $\vec{h}=h\hat{c}$. The KMF
denotes using Kitaev's decomposition to do MF. (c)Comparison between RMFT and VMC\cite{Wang_PRL} for calculated cone positions of the 14-cone QSL state of  $K$-$\Gamma$ model when $K=-1$ and $\Gamma=0.3$. (d) Comparison between RMFT and VMC\cite{Wang_PRL} of the Chern number evolution of 14-cone state under field $\vec{h}=h\hat{c}$ when $K=-1$ and $\Gamma=0.3$.}
\end{figure*}

\begin{figure*}[htbp]
\includegraphics[height=3.25in,width=5 in] {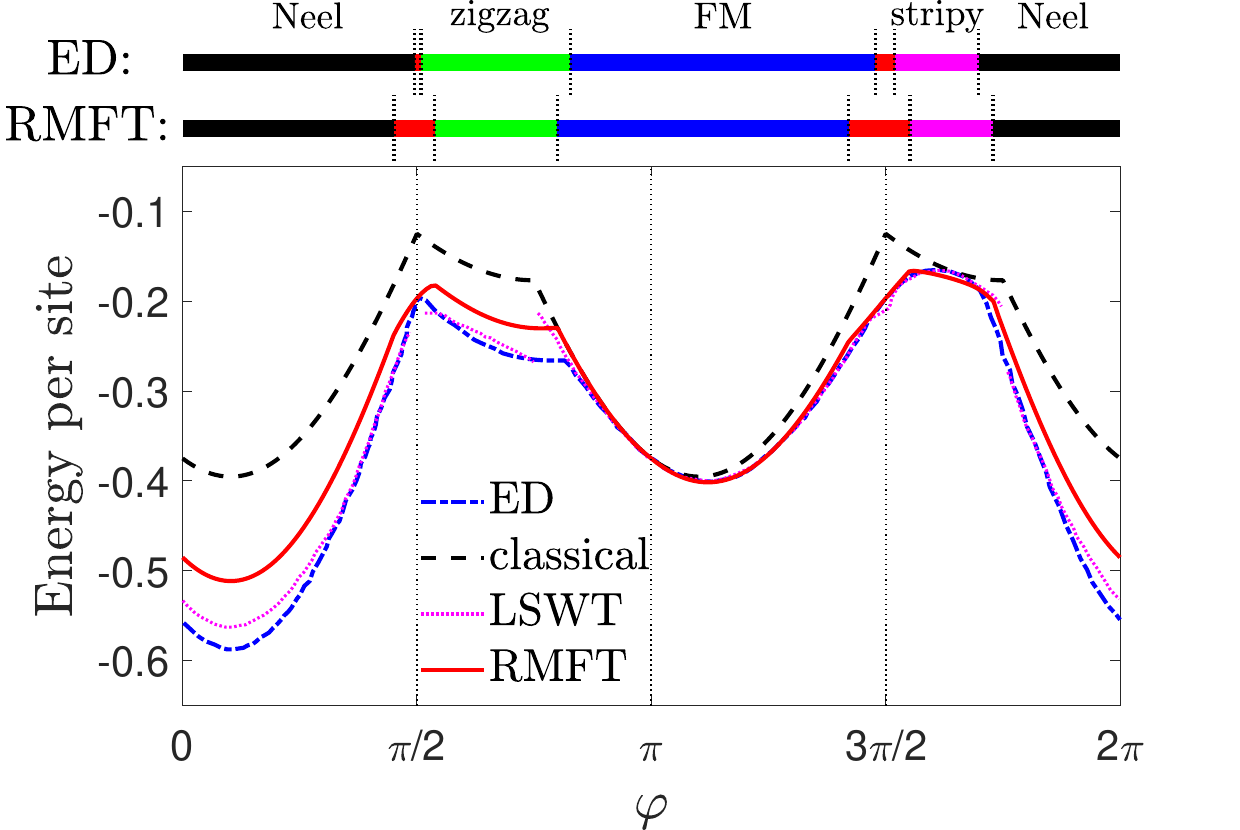}
\caption{ Comparison of RMFT results for the $J$-$K$ model\cite{KJmodel} at zero magnetic field, where the couplings are parameterized as
$J=\cos\varphi$,  $K=\sin\varphi$. Here "classical" denote the classical mean-field solution in which all $\chi_{ij}$'s are set to zero, "ED" represents exact diagonalization, and "LSWT" refers to linear spin wave theory.  Other numerical results presented in Ref. \cite{KJmodel} are close to the ED data and are therefore not shown here.}
\end{figure*}

\begin{figure*}[htbp]
\includegraphics[height=2.4in,width=6.2in] {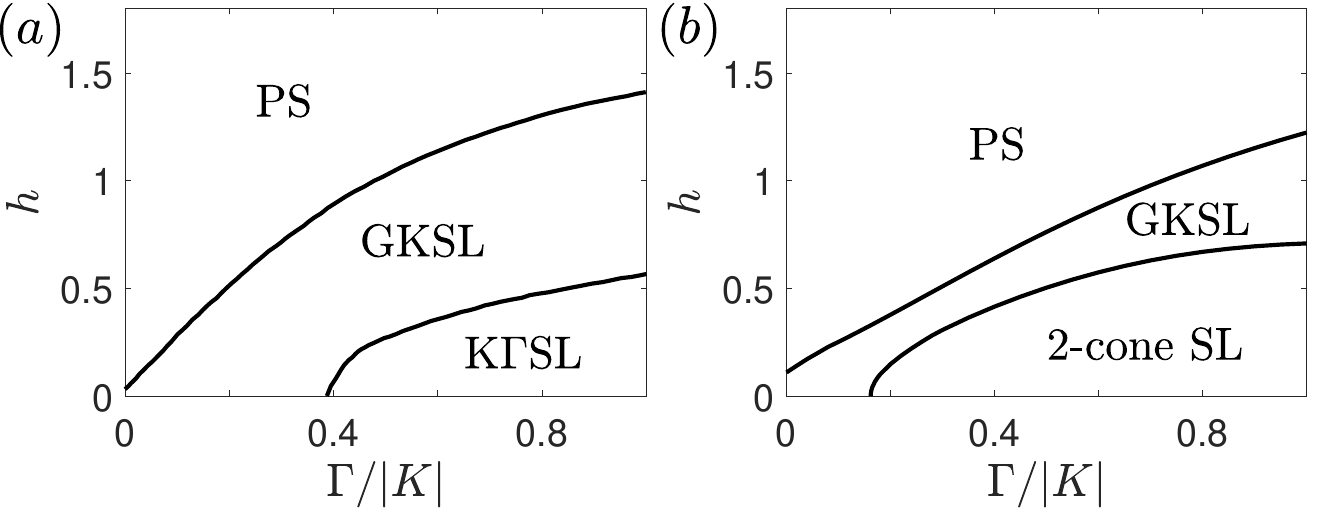}
\caption{Comparison of results for the $K$-$\Gamma$ model under applied field: (a) DMRG\cite{Kee4}. (b)RMFT. Here the couplings are parameterized as
 $K<0$, $K^2+\Gamma^2=1$, and the magnetic field is applied along the $\hat{c}$ axis ($\vec{h}=h\hat{c}$). "GKSL" denotes the generalized Kitaev spin-liquid state,
 and "PS" denotes the polarized state.}
\end{figure*}

\begin{figure*}[htbp]
\includegraphics[height=3.8in,width=6.6in] {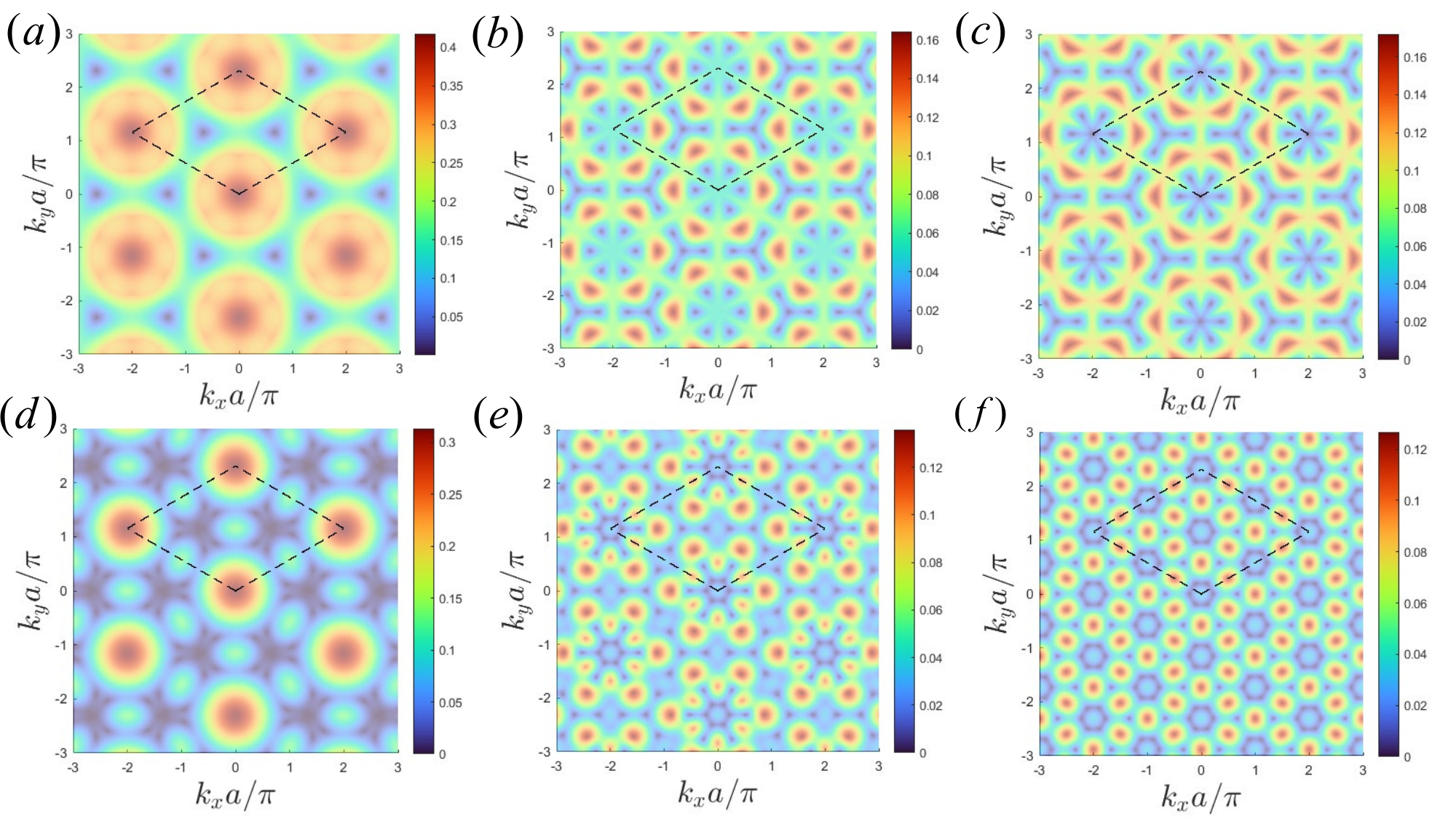}
\caption{Lowest fermionic excitation spectrum of different QSL states which survive for large $\Gamma/|K|$(up to 1): (a)2- (b)$14_2$- (c)16- (d)$20_2$- (e)$20_4$- (f)32-cone. Here $a$ is the lattice constant for the honeycomb lattice, the color scale represents the energy in unit of $|K|$, and the region enclosed by dash line is the first Brillouin zone. }
\end{figure*}

\begin{figure*}[htbp]
\includegraphics[height=1.9 in,width=7in] {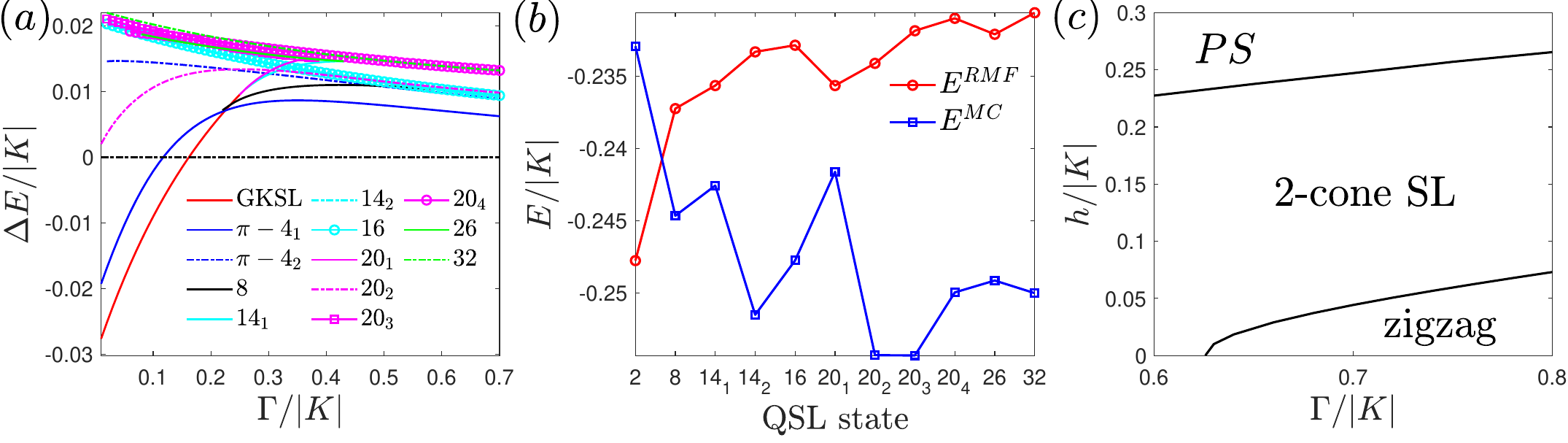}
\caption{Energy comparison of different quantum spin liquid (QSL) states and the resulting phase diagram for fixed $K=-1$: (a) Energy of various QSL states relative to the 2-cone state, plotted as a function of $\Gamma/K$, calculated using RMFT. Here $\Delta E = E-E_{2\mbox{-cone}}$. Note that $\pi-4_1$ state at $\Gamma=0$ corresponds to the full-vortex state in the Kitaev's exact solution.
(b)  Energy comparison of each QSL state computed using both RMFT and Monte Carlo (MC) methods at $\Gamma/|K|=0.3$.
(c) RMFT-calculated phase diagram for $\Gamma'=-0.08$, with a magnetic field applied along the $\hat{a}$ axis ($\vec{h}=h\hat{a}$). Here "zigzag" denotes the zigzag magnetically ordered state, and "PS" denotes the polarized state.}
\end{figure*}

\begin{figure*}[htbp]
\includegraphics[height=3.6in,width=3.4in] {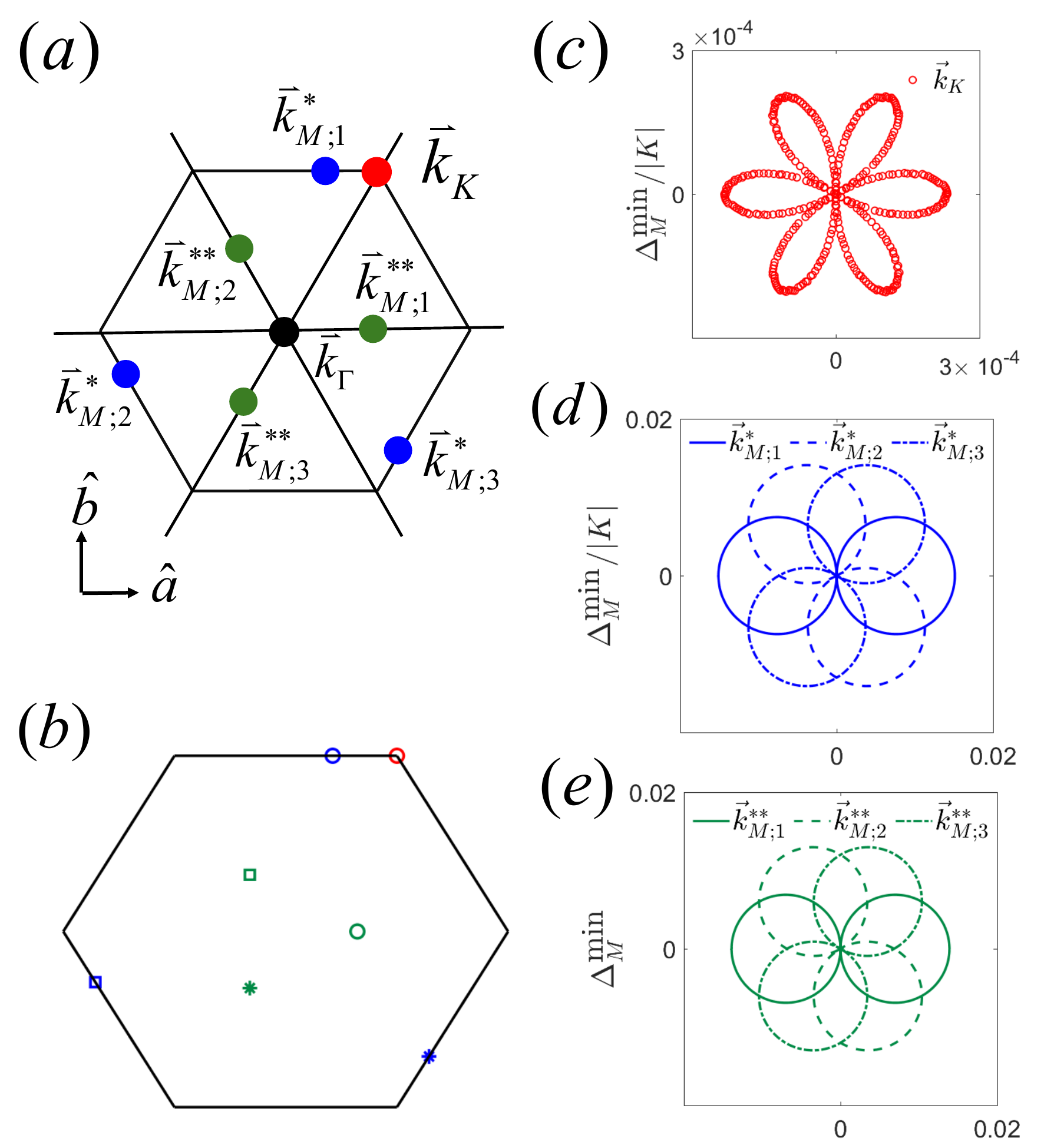}
\caption{Locations of Dirac cones in QSL states and angle dependence of gap: (a) Schematic plot of typical locations  of Dirac cones in momentum space for QSL states.
(b) Exact momentum positions of the Dirac cones from $14_2$-cone state under zero field,
where $\Gamma=0.3$. Note that cones at $-\vec{k}$ are omitted due to PHS. (c) Azimuthal angle-dependent minimum gap near $\vec{k}_K$.
(d) Azimuthal angle-dependent minimum gap near $\vec{k}^{*}_{M;1,2,3}$. (e) Azimuthal angle-dependent minimum gap near $\vec{k}^{**}_{M;1,2,3}$.
Here $\vec{h}=h(\cos\varphi\hat{a}+\sin\varphi\hat{b})$ and $h=0.01$.}
\end{figure*}
\begin{figure*}[htbp]
\includegraphics[height=1.5in,width=3.4in] {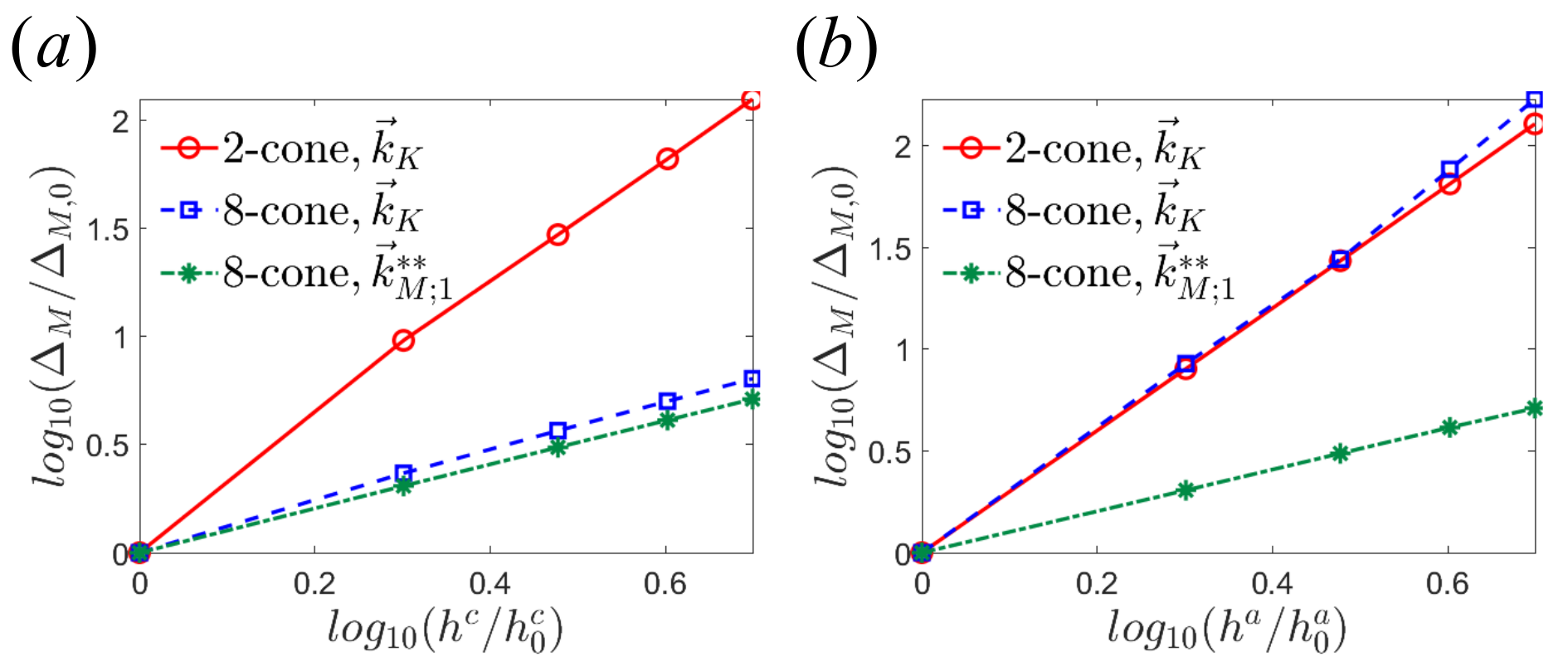}
\caption{Minimum gap near $\vec{k}_D$ from $2$-cone and $8$-cone states: (a) $\vec{h}\|\hat{c}$, (b) $\vec{h}\|\hat{a}$.
Here $\Gamma=0.5$, $h=0.01\sim0.05$, $h_0=0.01$, $\Delta_{M,0}$ is the gap value when $h=h_0$.}
\end{figure*}
\begin{figure*}[htbp]
\includegraphics[height=2.4in,width=3.55in] {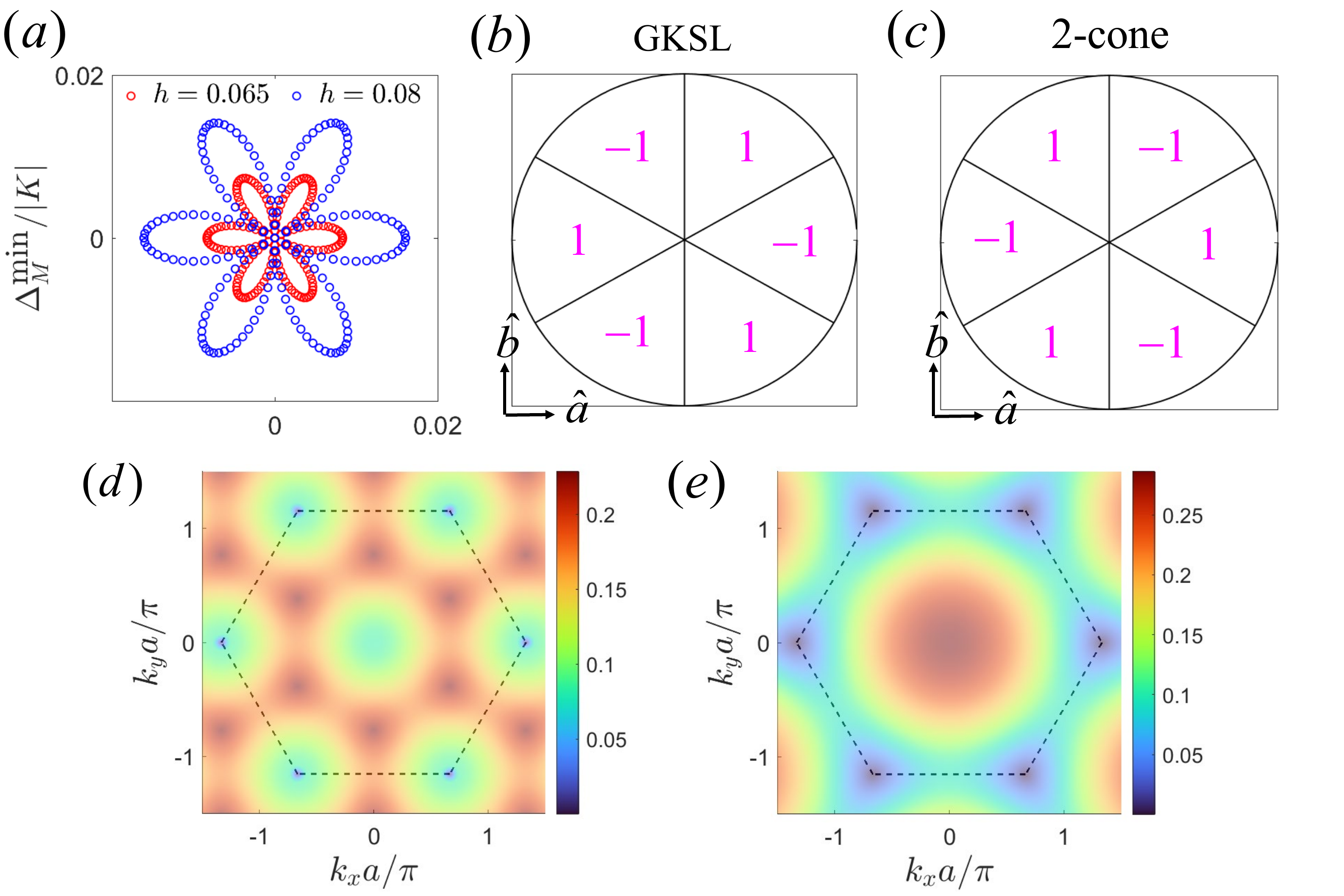}
\caption{Comparison of GKSL state with the found 2-cone state: (a) Fermionic gap of 2-cone state versus orientation of magnetic fields in the $\hat{a}$-$\hat{b}$ plane.
(b) Chern number of GKSL versus  orientation of magnetic fields in the $\hat{a}$-$\hat{b}$ plane. (c) Chern number of 2-cone state versus  orientation of magnetic fields in the $\hat{a}$-$\hat{b}$ plane. (d) Spectrum of GKSL at $\Gamma=0.16$. (e) Spectrum of 2-cone state at $\Gamma=0.16$. }
\end{figure*}
\begin{figure*}[htbp]
\includegraphics[height=3.6in,width=3.4in] {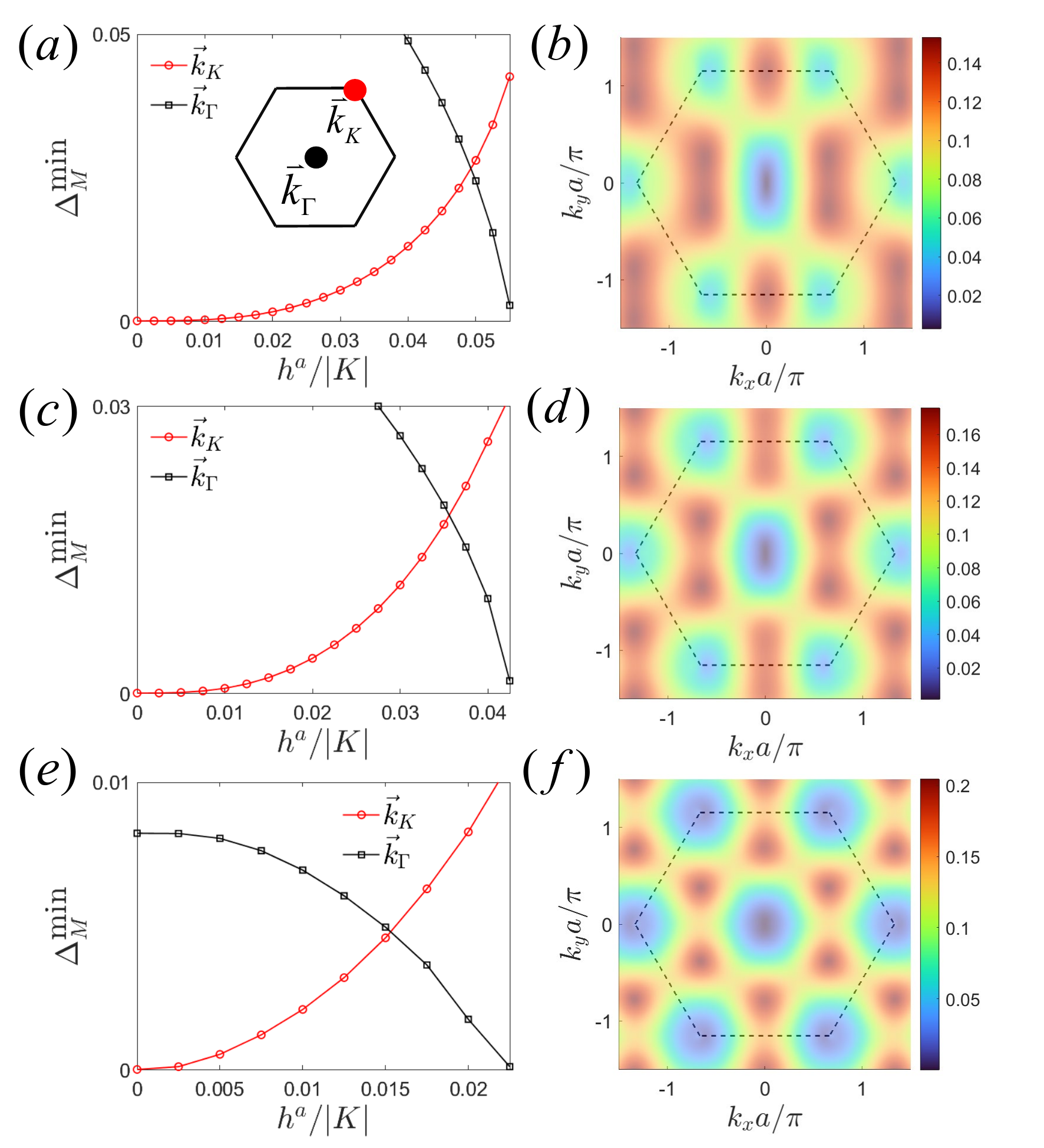}
\caption{Fermionic gap versus magnetic field $\vec{h}$ nearby $\vec{k}_K$ and $\vec{k}_{\Gamma}$ for GKSL state when $\vec{h}=h^a\hat{a}$: (a) \& (b) $\Gamma=0.16$ and the corresponding spinon spectrum near the critical field when $\Delta^{min}_M \sim 0$ at $\vec{k}_{\Gamma}$. (c) \& (d) $\Gamma=0.2$ and  the corresponding spinon spectrum near the critical field. (e) \& (f) $\Gamma=0.24$ and the corresponding spinon spectrum near the critical field.}
\end{figure*}
\begin{figure*}[htbp]
\includegraphics[height=3.6in,width=3.4in] {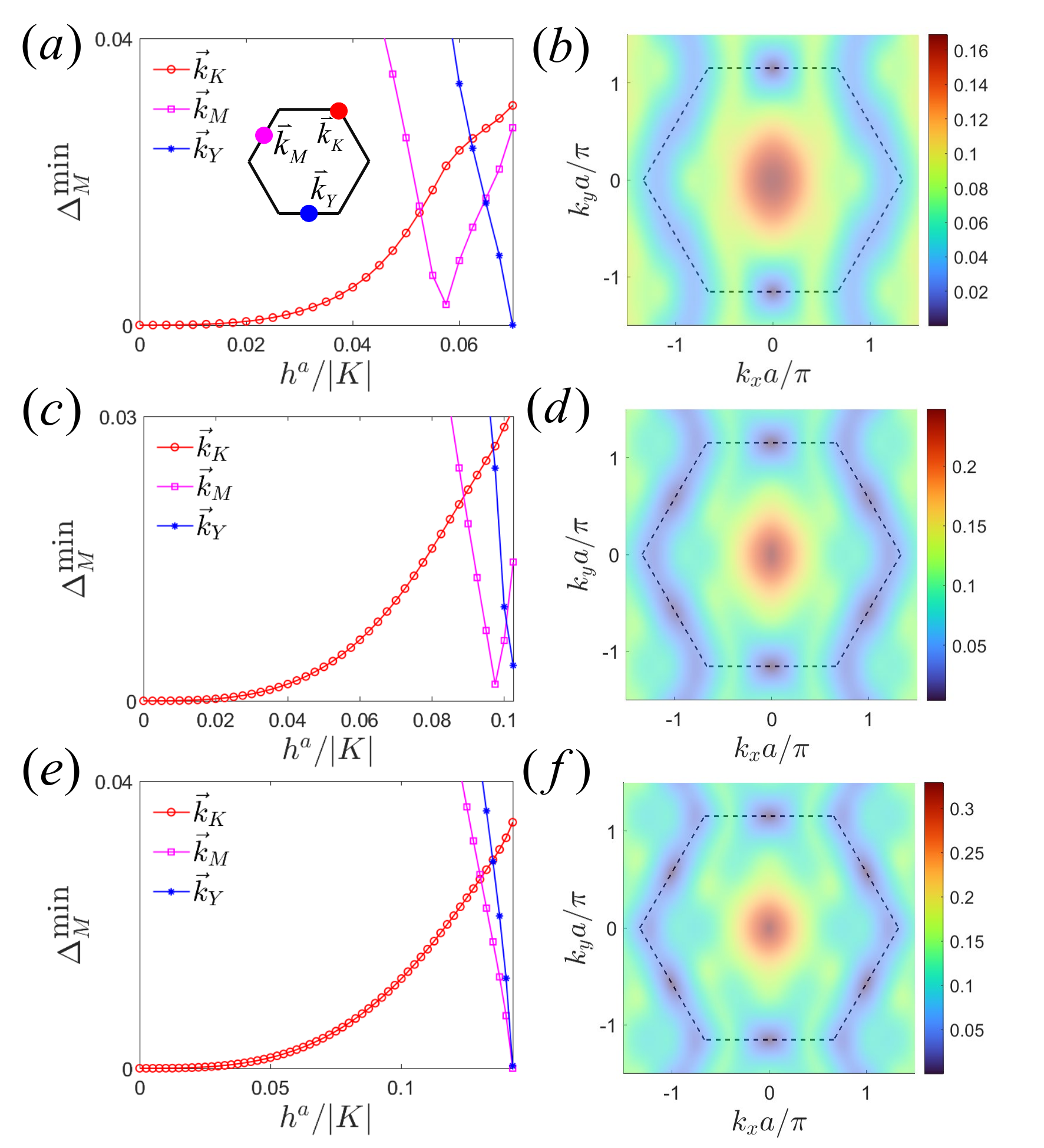}
\caption{Fermionic gap versus magnetic field $\vec{h}$ nearby $\vec{k}_K$ and $\vec{k}_{\Gamma}$ for the found 2-cone state when $\vec{h}=h^a\hat{a}$: (a) \& (b) $\Gamma=0.16$ and the corresponding spinon spectrum near the critical field when $\Delta^{min}_M \sim 0$ at $\vec{k}_M$ and $\vec{k}_Y$.  (c) \& (d) $\Gamma=0.3$ and the corresponding spinon spectrum near the critical field. (e) \& (f) $\Gamma=0.5$ and the corresponding spinon spectrum near the critical field.}
\end{figure*}
\begin{figure*}[htbp]
\includegraphics[height=1.6in,width=3.65in] {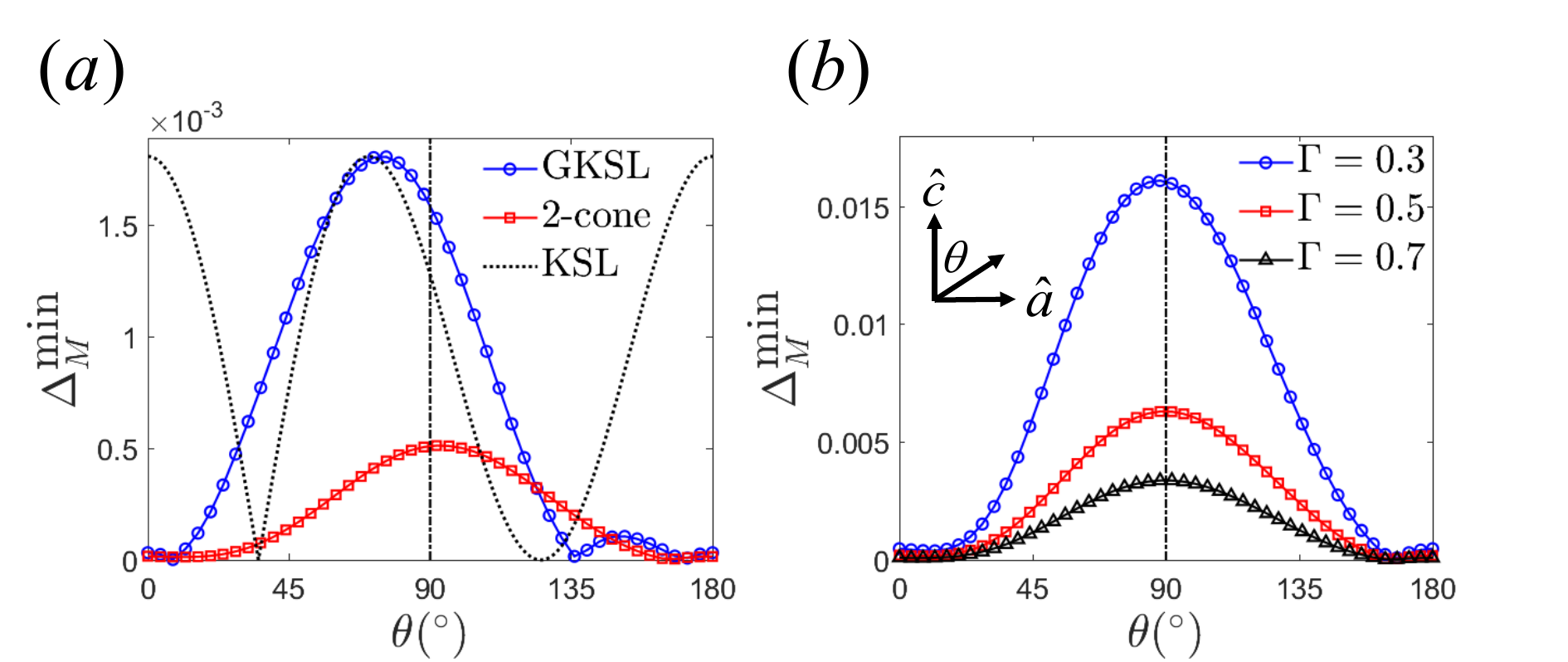}
\caption{Comparison of field polar-angle dependence of the fermionic gap: (a) the GKSL state and the 2-cone state with $\Gamma=0.16$ and $h=0.02$,
where the black dot line denotes the schematic fermionic gap of the KSL state, which is proportional to $|h^xh^yh^z|$ under the field.
 (b) 2-cone state with different $\Gamma$ at $h=0.08$. }
\end{figure*}
\begin{figure*}[htbp]
\includegraphics[height=4.8in,width=3.4in] {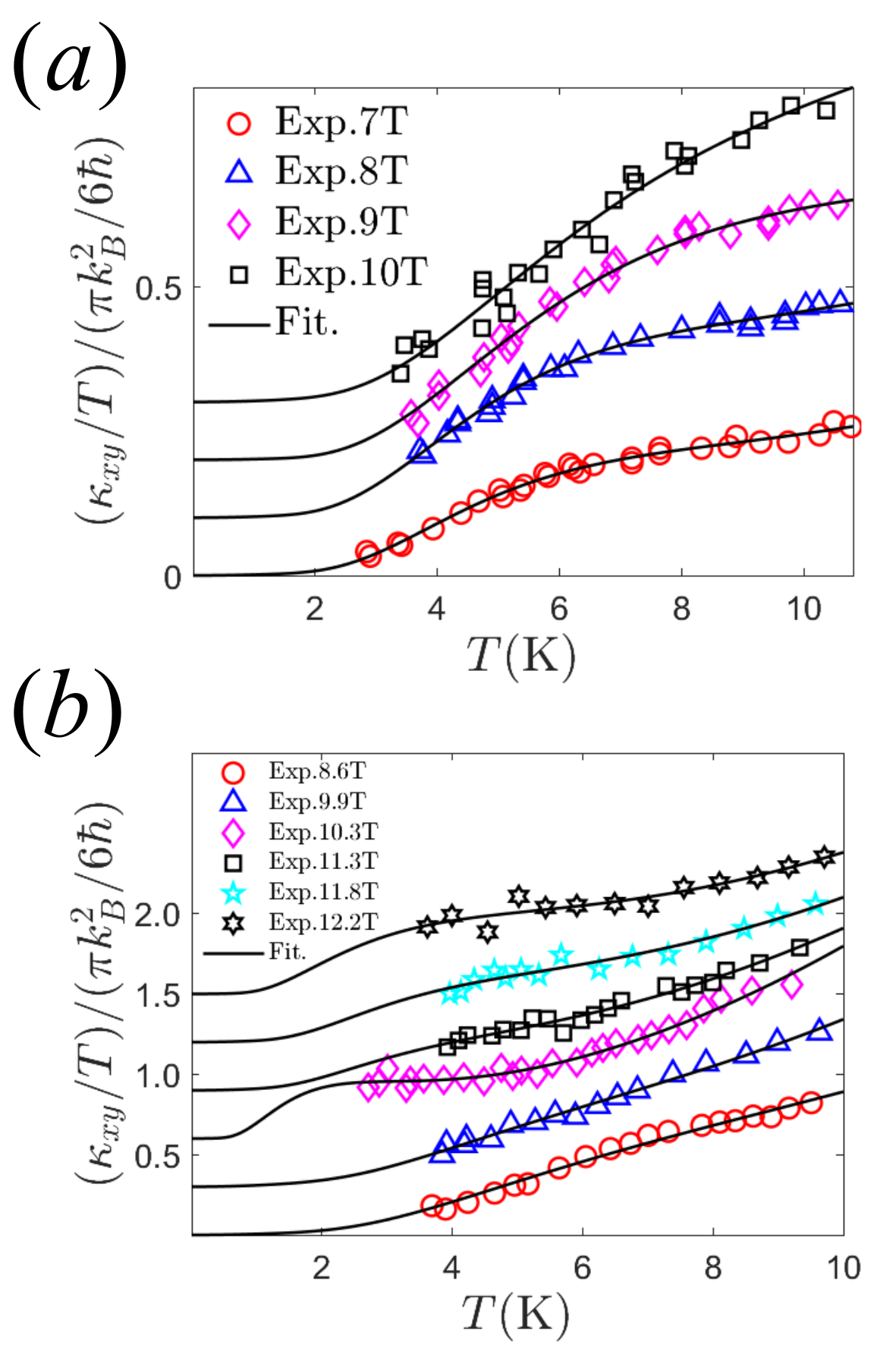}
\caption{ Fitting the experimental data by using gapped edge state. Here we fit the data by using the expression $(\kappa_{xy}/T)/(\pi k^2_B/6\hbar)=F(\Lambda/k_BT) -  F(\Delta/k_BT)+\alpha T^2$. Experimental data taken from (a) Czajka's group \cite{Czajka_2}, (b) Bruin's group \cite{Bruin}.
To exhibit data points and the curves clearly, data points and curves are shifted by 0.1 in y-axis successively for each $B$ field in (a),
and are shifted by 0.3 for each $B$ field in (b). The detail of fitting parameter is shown in sec. X of the Supplemental Material\cite{SMI}.}
\end{figure*}

\FloatBarrier

\begin{table*}[htbp] \label{table1}
\resizebox{0.35\textwidth}{!}{
 \begin{tabular}{|c|c|c|}
 \hline
 QSL state & $E^{\mbox{\tiny{MC}}}/E^{\mbox{\tiny{RMF}}}$ & $W_{\mbox{\tiny{MC}}}$   \\ [0.5ex]
 \hline
 $\mbox{GKSL}$,$\Gamma=0.05$  & $0.996\pm0.008$ & $0.902\pm0.001$ \\
 $\mbox{GKSL}$,$\Gamma=0.10$  & $0.997\pm0.006$ & $0.848\pm0.002$ \\
 $\mbox{GKSL}$,$\Gamma=0.15$  & $0.996\pm0.009$ & $0.794\pm0.005$ \\
 $\mbox{GKSL}$,$\Gamma=0.20$  & $1.000\pm0.006$ & $0.729\pm0.002$ \\
 \hline
\end{tabular}}
\caption{Comparison of RMFT and MC results, both of which are calculated on $7\times7\times2$ lattice.
The error of ratio $E^{\mbox{\tiny{MC}}}/E^{\mbox{\tiny{RMF}}}$ for GKSL is only about 1\%.}
\end{table*}

\begin{table*}[htbp] \label{table2}
\resizebox{0.33\textwidth}{!}{
 \begin{tabular}{|c|c|c|}
 \hline
 QSL state & $E^{\mbox{\tiny{MC}}}/E^{\mbox{\tiny{RMF}}}$ & $W_{\mbox{\tiny{MC}}}$   \\ [0.5ex]
 \hline
 $2\mbox{-cone}$    & $0.940\pm0.003$ & $-0.353\pm0.003$ \\
 $8\mbox{-cone}$    & $1.031\pm0.004$ & $-0.034\pm0.004$ \\
 $14_1\mbox{-cone}$ & $1.029\pm0.005$ & $0.279\pm0.004$ \\
 $14_2\mbox{-cone}$ & $1.078\pm0.008$ & $-0.188\pm0.005$ \\
 $16\mbox{-cone}$   & $1.064\pm0.004$ & $-0.187\pm0.002$ \\
 $20_1\mbox{-cone}$ & $1.025\pm0.006$ & $0.280\pm0.006$ \\
 $20_2\mbox{-cone}$ & $1.086\pm0.006$ & $-0.209\pm0.004$ \\
 $20_3\mbox{-cone}$ & $1.097\pm0.007$ & $-0.161\pm0.004$ \\
 $20_4\mbox{-cone}$ & $1.082\pm0.011$ & $-0.156\pm0.005$ \\
 $26\mbox{-cone}$   & $1.073\pm0.014$ & $-0.220\pm0.004$ \\
 $32\mbox{-cone}$   & $1.084\pm0.015$ & $-0.218\pm0.003$ \\
 \hline
\end{tabular}}
\caption{Comparison of RMFT and MC results, both of which are calculated on $7\times7\times2$ lattice with $\Gamma=0.3$.}
\end{table*}


\begin{thebibliography}{99}

\bibitem{Balent} Savary, L. \& Balents, L. Quantum Spin Liquids: a review. \textit{Rep. Prog. Phys.} {\bf 80}, 016502 (2017).
\bibitem{Anderson1} Anderson, P. W. Resonating valence bonds: A new kind of insulator? \textit{Materials Research Bulletin} {\bf 8}, 153 (1973).
\bibitem{Anderson2} Anderson, P. W.  The Resonating Valence Bond State in La$_2$CuO$_4$ and Superconductivity. \textit{Science} {\bf 235}, 1196 (1987).

\bibitem{Kitaev1}  Kitaev, A. Anyons in an exactly solved model and beyond. \textit{Annals of Physics} {\bf 321}, 2-111 (2006).
\bibitem{Nasu1} Nasu, J., Kato, Y., Kamiya, Y., \& Motome, Y. Successive Majorana topological transitions driven by a magnetic field in the Kitaev model. \textit{Phys. Rev. B} {\bf 98}, 060416(R) (2018).
\bibitem{Gohlke1} Gohlke, M., Moessner, R. \& Pollmann, F. Dynamical and topological properties of the Kitaev model in a [111] magnetic field. \textit{Phys. Rev. B} {\bf 98}, 014418 (2018).
\bibitem{Zhang} Zhang, S.-S., Halász, G. B. \& Batista, C. D. Theory of the Kitaev model in a [111] magnetic field. \textit{Nat. Commun.} {\bf 13}, 399 (2022).
\bibitem{Hickey} Hickey, C. \& Trebst, S. Emergence of a field-driven U(1) spin liquid in the Kitaev honeycomb model. \textit{Nat. Commun.} {\bf 10}, 530 (2019).

\bibitem{Nasu2} Nasu, J., Yoshitake, J. \& Motome, Y. Thermal Transport in the Kitaev Model. \textit{Phys. Rev. Lett.} {\bf 119}, 127204 (2017).
\bibitem{Matsuda_Nat} Kasahara, Y. et al. Majorana quantization and half-integer thermal quantum Hall effect in a Kitaev spin liquid.
                      \textit{Nature.} {\bf 559}, 227–231 (2018).
\bibitem{Matsuda_Sci} Yokoi, T. et al. Half-integer quantized anomalous thermal Hall effect in the Kitaev material candidate $\alpha$-RuCl$_3$.
                      \textit{Science.} {\bf 373}, 568–572 (2021).
\bibitem{Matsuda_PRB} Kasahara, Y. et al. Quantized and unquantized thermal Hall conductance of the Kitaev spin liquid candidate $\alpha$-RuCl$_3$.
                      \textit{Phys. Rev. B} {\bf 106}, L060410 (2022).

\bibitem{compass1} Jackeli, G. \& Khaliullin, G. Mott Insulators in the Strong Spin-Orbit Coupling Limit: From Heisenberg to a Quantum Compass and Kitaev Models.
                     \textit{Phys. Rev. Lett.} {\bf 102}, 017205 (2009).
\bibitem{compass2} Chaloupka, J., Jackeli, G. \& Khaliullin, G. Kitaev-Heisenberg Model on a Honeycomb Lattice: Possible Exotic Phases in Iridium Oxides A$_2$IrO$_3$.
                    \textit{Phys. Rev. Lett.} {\bf 105}, 027204 (2010).

\bibitem{Katukuri} Katukuri, V. M. et al. Kitaev interactions between $j = 1/2$ moments in honeycomb Na$_2$IrO$_3$ are large and ferromagnetic: insights from ab initio quantum chemistry calculations. \textit{New J. Phys.} {\bf 16}, 013056 (2014).
\bibitem{Kee2} Rau, J. G., Lee, E. K.-H. \& Kee, H.-Y. Generic Spin Model for the Honeycomb Iridates beyond the Kitaev Limit. \textit{Phys. Rev. Lett.} {\bf 112}, 077204 (2014).
\bibitem{Yamaji}  Yamaji, Y., Nomura, Y., Kurita, M., Arita, R. \& Imada, M.
                  First-Principles Study of the Honeycomb-Lattice Iridates Na$_2$IrO$_3$ in the Presence of Strong Spin-Orbit Interaction and Electron
                  Correlations. \textit{Phys. Rev. Lett.} {\bf 113}, 107201 (2014).
\bibitem{Kee3} Rau, J. G., \& Kee, H.-Y. Trigonal distortion in the honeycomb iridates: Proximity of zigzag and spiral phases in Na$_2$IrO$_3$. \textit{ArXiv}: 1408.4811v1 (2014).
\bibitem{Janssen}  Janssen, L., Andrade, E. C. \&  Vojta, M. Magnetization processes of zigzag states on the honeycomb lattice: Identifying spin models for $\alpha$-RuCl$_3$ and Na$_2$IrO$_3$. \textit{Phys. Rev. B} {\bf 96}, 064430 (2017).
\bibitem{Pontus} Laurell, P. \& Okamoto, S. Dynamical and thermal magnetic properties of the Kitaev spin liquid candidate $\alpha$-RuCl$_3$.
                 \textit{npj Quantum Materials} {\bf 5}, 2 (2020).

\bibitem{Takagi} Takagi, H., Takayama, T., Jackeli, G., Khaliullin, G. \& Nagler, S. E.
                  Concept and realization of Kitaev quantum spin liquids. \textit{Nature Reviews Physics} {\bf 1}, 264 (2019).
\bibitem{Chaebin} Kim, C. et al. Antiferromagnetic Kitaev interaction in J$_{\mbox{eff}}$ = 1/2 cobalt honeycomb materials Na$_3$Co$_2$SbO$_6$ and Na$_2$Co$_2$TeO$_6$.
                  \textit{Journal of Physics: Condensed Matter} {\bf 34}, 045802 (2021).
\bibitem{Lin} Lin, G. et al. Field-induced quantum spin disordered state in spin-1/2 honeycomb magnet Na$_2$Co$_2$TeO$_6$. \textit{Nat. Commun.} {\bf 12}, 5559 (2021).
\bibitem{Weiliang} Yao, W., Iida, K., Kamazawa, K. \& Li, Y.
                    Excitations in the Ordered and Paramagnetic States of Honeycomb Magnet Na$_2$Co$_2$TeO$_6$. \textit{Phys. Rev. Lett.} {\bf 129}, 147202 (2022).

\bibitem{Czajka_1} Czajka, P. et al. Oscillations of the thermal conductivity in the spin-liquid state of $\alpha$-RuCl$_3$. \textit{Nat. Phys.} {\bf 17}, 915–919 (2021).
\bibitem{Bruin} Bruin, J. A. N. et al. Robustness of the thermal Hall effect close to half-quantization in $\alpha$-RuCl$_3$. \textit{Nat. Phys.} {\bf 18}, 401–405 (2022).
\bibitem{Czajka_2} Czajka, P. et al. Planar thermal Hall effect of topological bosons in the Kitaev magnet $\alpha$-RuCl$_3$. \textit{Nat. Mater.} {\bf 22}, 36–41 (2023).
\bibitem{Kee1} Kee, H.-Y. Thermal Hall conductivity of $\alpha$-RuCl$_3$. \textit{Nat. Mater.} {\bf 22}, 6–7 (2023).
\bibitem{Lefran} Lefrançois, É. et al. Evidence of a Phonon Hall Effect in the Kitaev Spin Liquid Candidate $\alpha$-RuCl$_3$. \textit{Phys. Rev. X} {\bf 12}, 021025 (2022).

\bibitem{Tanaka} Tanaka, O. et al. Thermodynamic evidence for a field-angle-dependent Majorana gap in a Kitaev spin liquid. \textit{Nat. Phys.} {\bf 18}, 429–435 (2022)
\bibitem{Imamura} Imamura, K. et al. Majorana-fermion origin of the planar thermal Hall effect in the Kitaev magnet $\alpha$-RuCl$_3$. \textit{Sci. Adv.} 10, eadk3539 (2024).
\bibitem{Hwang} Hwang, K., Go, A., Seong, J-.H., Shibauchi, T. \& Moon, E.-G.
                Identification of a Kitaev quantum spin liquid by magnetic field angle dependence.  \textit{Nat. Commun.} {\bf 13}, 323 (2022).

\bibitem{Gohlke2} Gohlke, M., Wachtel, G., Yamaji, Y., Pollmann, F. \& Kim, Y. B.
                  Quantum spin liquid signatures in Kitaev-like frustrated magnets. \textit{Phys. Rev. B} {\bf 97}, 075126 (2018).
\bibitem{Kee4} Gordon, J. S., Catuneanu, A., Sørensen, E. S. \&  Kee, H.-Y. Theory of the field-revealed Kitaev spin liquid.  \textit{Nat. Commun.} {\bf 10}, 2470 (2019).
\bibitem{Lee} Lee, H.-Y. et al. Magnetic field induced quantum phases in a tensor network study of Kitaev magnets. \textit{Nat. Commun.} {\bf 11}, 1639 (2020).
\bibitem{toronto} Luo, Q., Zhao, J., Kee, H.-Y. \& Wang, X. Gapless quantum spin liquid in a honeycomb $\Gamma$ magnet, \textit{npj Quantum Materials} {\bf 6}, 57 (2021).
\bibitem{Yilmaz} Yılmaz, F., Kampf, A. P. \& Yip, S. K. Phase diagrams of Kitaev models for arbitrary magnetic field orientations. \textit{Phys. Rev. Res.} {\bf 4}, 043024 (2022).

\bibitem{PSG_Wen} Wen, X.-G. Quantum orders and symmetric spin liquids. \textit{Phys. Rev. B} {\bf 65}, 165113 (2002).
\bibitem{PSG_Burnell} Burnell, F. J. \& Nayak, C. SU(2) slave fermion solution of the Kitaev honeycomb lattice model. \textit{Phys. Rev. B} {\bf 84}, 125125 (2011).
\bibitem{PSG_You} You, Y.-Z., Kimchi, I. \&  Vishwanath, A. Doping a spin-orbit Mott insulator: Topological superconductivity from the Kitaev-Heisenberg model and possible application to (Na$_2$/Li$_2$)IrO$_3$. \textit{Phys. Rev. B} {\bf 86}, 085145 (2012).

\bibitem{Wang_PRL} Wang, J.,  Normand, B. \& Liu, Z.-X. One Proximate Kitaev Spin Liquid in the K-J-$\Gamma$ Model on the Honeycomb Lattice. \textit{Phys. Rev. Lett.} {\bf 123}, 197201 (2019).

\bibitem{Wen_Lee} Lee, P. A., Nagaosa, N. \&  Wen, X.-G. Doping a Mott insulator: Physics of high-temperature superconductivity. \textit{Rev. Mod. Phys.} {\bf 78}, 17 (2006).

\bibitem{Wang_PRB} Wang, J., Zhao, Q., Wang, X. \& Liu, Z.-X.  Multinode quantum spin liquids on the honeycomb lattice. \textit{Phys. Rev. B} {\bf 102}, 144427 (2020).
\bibitem{Jiang_PRL} Jiang, M.-H. et al. Tuning Topological Orders by a Conical Magnetic Field in the Kitaev Model.  \textit{Phys. Rev. Lett.} {\bf 125}, 177203 (2020).

\bibitem{KJmodel}Gotfryd, D., Rusnačko, J., Wohlfeld, K., Jackeli, G., Chaloupka, J., \&  Oleś, A. M. Phase diagram and spin correlations of the Kitaev-Heisenberg model: Importance of quantum effects. \textit{Phys. Rev. B} {\bf 95}, 024426 (2017).

\bibitem{MajDecom1} Chen, G., Essin, A. \& Hermele, M. Majorana spin liquids and projective realization of SU(2) spin symmetry. \textit{Phys. Rev. B} {\bf 85}, 094418 (2012).
\bibitem{MajDecom2} Fu, J., Knolle, J., \& Perkins, N. B. Three types of representation of spin in terms of Majorana fermions and an alternative
solution of the Kitaev honeycomb model. \textit{Phys. Rev. B} {\bf 97}, 115142 (2018).

\bibitem{RMFT1} Zhang, F. C., Gros, C., Rice, T. M. \&  Shiba, H. A renormalised Hamiltonian approach to a resonant valence bond wavefunction. \textit{Supercond. Sci. Technol.} {\bf 1}, 36 (1988).
\bibitem{RMFT2} Edegger, B., Muthukumar, V. N. \& Gros, C.  Gutzwiller–RVB theory of high-temperature superconductivity: Results from renormalized mean-field theory and variational Monte Carlo calculations. \textit{Adv. Phys.} {\bf 56}, 927 (2007).
\bibitem{RMFT3} Ogata, M. \&  Himeda, A. Superconductivity and Antiferromagnetism in an Extended Gutzwiller Approximation for t–J Model: Effect of Double-Occupancy Exclusion. \textit{J. Phys. Soc. Jpn.} {\bf 72}, 374 (2003).
\bibitem{RMFT4} Anderson,  P. W. et al. The physics behind high-temperature superconducting cuprates: the 'plain vanilla' version of RVB. \textit{J. Phys.: Condens. Matter} {\bf 16 }, R755 (2004).
\bibitem{RMFT5} Fukushima, N.  Grand canonical Gutzwiller approximation for magnetic inhomogeneous systems. \textit{Phys. Rev. B} {\bf 78}, 115105(2008)
\bibitem{RMFT6} Choubey, P., Tu, W. L., Lee, T. K. \& Hirschfeld, P. J. Incommensurate charge ordered states in the t–t'–J model. \textit{New J. Phys.} {\bf 19}, 013028(2017).
\bibitem{RMFT_compare} Zhu, Z., Kimchi, I., Sheng, D. N. \& Fu, L. Robust non-Abelian spin liquid and a possible intermediate phase in the antiferromagnetic Kitaev model with magnetic field. \textit{Phys. Rev. B} {\bf 97}, 241110 (2018).


\bibitem{Perkin} Rousochatzakis, I., Perkins, N.B., Luo, Q. \& Kee, H.-Y. Beyond Kitaev physics in strong spin-orbit coupled magnets. \textit{Reports on Progress in
Physics.} {\bf 87}, 026502 (2024).

\bibitem{TP_Wen} Wen, X.-G. Colloquium: Zoo of quantum-topological phases of matter. \textit{Rev. Mod. Phys.} {\bf 89}, 041004 (2017).

\bibitem{KitaevPhy1} Pedrocchi, F. L.,  Chesi, S. \& Loss, D. Physical solutions of the Kitaev honeycomb model. \textit{Phys. Rev. B} {\bf 84}, 165414 (2011).
\bibitem{KitaevPhy2} Zschocke, F.  \&  Vojta, M. Physical states and finite-size effects in Kitaev's honeycomb model: Bond disorder, spin excitations, and NMR line shape. \textit{Phys. Rev. B} {\bf 92}, 014403 (2015).

\bibitem{RMFT7} Tu, W.-L., Schindler, F., Neupert, T. \& Poilblanc, D. Competing orders in the Hofstadter t-J model. \textit{Phys. Rev. B} {\bf 97}, 035154 (2018).

\bibitem{Suetsugu} Suetsugu, S. et al. Evidence for a Phase Transition in the Quantum Spin Liquid State of a Kitaev Candidate $\alpha$-RuCl$_3$. \textit{J. Phys. Soc. Jpn.} {\bf 91}, 124703 (2022).

\bibitem{KStrength_1}Do, S.-H. et al. Majorana fermions in the Kitaev quantum spin system. \textit{Nat. Phys.} {\bf 13}, 1079 (2017).
\bibitem{KStrength_2}Laurell, P. \&  Okamoto, S. Dynamical and thermal magnetic properties of the Kitaev spin liquid candidate $\alpha$-RuCl$_3$. \textit{npj Quantum Mater.} {\bf 5}, 2 (2020).

\bibitem{WeiLi} Li, H. et al. Identification of magnetic interactions and high-field quantum spin liquid in $\alpha$-RuCl$_3$. \textit{Nat. Commun.} {\bf 12}, 4007(2021).

\bibitem{PhononHall1} Ye, M., Halász, G.B., Savary, L. \& Balents, L. Quantization of the Thermal Hall Conductivity at Small Hall Angles. \textit{Phys. Rev. Lett.} {\bf 121}, 147201 (2018).
\bibitem{PhononHall2} Aviv, Y.V. \& Rosch, A. Approximately Quantized Thermal Hall Effect of Chiral Liquids Coupled to Phonons. \textit{Phys. Rev. X }{\bf 8}, 031032 (2018).

\bibitem{PhononHall3} Qin, T., Zhou, J. \& Shi, J. Berry curvature and the phonon Hall effect. \textit{Phys. Rev. B} {\bf 86}, 104305 (2012).
\bibitem{PhononHall4} Chen, J.-Y., Kivelson, S. A. \& Sun, X.-Q. Enhanced Thermal Hall Effect in Nearly Ferroelectric Insulators. \textit{Phys. Rev. Lett.} {\bf 124}, 167601 (2020).
\bibitem{PhononHall5} Yip, S.-K. Coupling of acoustic phonons to spin-orbit entangled pseudospins. \textit{Phys. Rev. B} {\bf 108}, 165116 (2023).

\bibitem{SMI}
See Supplemental Material for details,
including the gauge invariant Majorana fermions representation and transformation,
full derivation of RMFT, PSG, IGG, and symmetry analysis of low energy effective model under field.
\end{thebibliography}
\end{document}